\documentclass{SciPost}

\hypersetup{
    colorlinks,
    linkcolor={red!50!black},
    citecolor={blue!50!black},
    urlcolor={blue!80!black}
}

\usepackage[bitstream-charter]{mathdesign}
\usepackage{graphicx}
\usepackage{amsmath}
\usepackage{color}
\usepackage{xpatch}
\usepackage{caption}
\usepackage{subcaption}
\usepackage{slashed}
\usepackage[capitalize]{cleveref}
\usepackage{tikz}
\usepackage{makecell}
\usetikzlibrary{decorations.pathreplacing}
\usepackage[normalem]{ulem}
\usepackage[T1]{fontenc}
\usepackage[htt]{hyphenat}

\usepackage{tabularray}
\usepackage{cprotect}

\DeclareSymbolFont{usualmathcal}{OMS}{cmsy}{m}{n}
\DeclareSymbolFontAlphabet{\mathcal}{usualmathcal}

\fancypagestyle{SPstyle}{
\fancyhf{}
\lhead{\colorbox{scipostblue}{\bf \color{white} ~SciPost Physics }}
\rhead{{\bf \color{scipostdeepblue} ~Submission }}

\fancyfoot[C]{\textbf{\thepage}}
}
\allowdisplaybreaks

\renewcommand{\arraystretch}{1.4}

\def\cA{\mathcal{A}}

\def\cM{\mathcal{M}}
\def\cF{\mathcal{F}}
\def\cR{\mathcal{R}}
\def\cH{\mathcal{H}}
\def\cO{\mathcal{O}}

\def\eps{\epsilon}

\DeclareMathOperator{\tr}{\rm tr}

\def\trm{\tr_-}
\def\trp{\tr_+}

\def\as{\alpha_s}
\def\zz{\mathbf{Z}}
\def\ll{\ell}

\def\bbh{b\bar{b}H}
\def\bbggh{\bar{b}bggH}
\def\bbqqh{\bar{b}b\bar{q}qH}

\def\bbggH{0\to\bar{b}bggH}
\def\bbqqH{0\to\bar{b}b\bar{q}qH}
\def\bbbbH{0\to\bar{b}b\bar{b}bH}

\def\lc{\mathrm{LC}}

\def\htl{\mathrm{HTL}}
\def\ren{\mathrm{ren}}

\def\la{\langle}
\def\ra{\rangle}
\def\spA#1#2{\la#1#2\ra}
\def\spB#1#2{[#1#2]}
\def\spAB#1#2#3{\la#1|#2|#3]}

\def\lo{\mathrm{LO}}

\renewcommand{\i}{\ensuremath{\mathrm{i}}}

\def\colhelsum{\underset{\mathrm{colour}}{\overline{\sum}}\,\underset{\mathrm{helicity}}{\overline{\sum}}}

\begin{document}

\pagestyle{SPstyle}

\begin{flushright}
  \footnotesize
  IFJPAN-IV-2026-8
\end{flushright}

\begin{center}{\Large \textbf{\color{scipostdeepblue}{
Top-Yukawa contributions to $pp\to b\bar{b}H$: two-loop leading-colour amplitudes\\
}}}\end{center}

\begin{center}\textbf{
Heribertus Bayu Hartanto\textsuperscript{1,2$\star$},
Rene Poncelet\textsuperscript{3$\dagger$}
}\end{center}

\begin{center}
{\bf 1} Asia Pacific Center for Theoretical Physics, Pohang, 37673, Korea \\
{\bf 2} Pohang University of Science and Technology, Pohang, 37673, Korea \\
{\bf 3} The Henryk Niewodnicza\'nski Institute of Nuclear Physics, ul.\ Radzikowskiego 152, 31-342 Krakow, Poland
\\[\baselineskip]
$\star$ \href{mailto:email1}{\small bayu.hartanto@apctp.org},
$\dagger$ \href{mailto:email1}{\small rene.poncelet@ifj.edu.pl}
\end{center}

\section*{\color{scipostdeepblue}{Abstract}}
\textbf{\boldmath{%
We derive two-loop scattering amplitudes for bottom-quark pair production in association with a Higgs boson at the LHC, 
focusing on terms proportional to the top-quark Yukawa coupling. We treat the bottom quark as a massless parton 
and employ both the leading-colour and heavy-top-quark approximations. 
The finite remainder of the two-loop amplitude is expressed in terms of one-mass pentagon functions, 
and the corresponding rational coefficients are reconstructed analytically from evaluations over finite fields. 
The scattering processes considered in this work also constitute a subset of Higgs+2-jet production at the LHC in the heavy-top-quark approximation.
}}

\vspace{\baselineskip}

\noindent\textcolor{white!90!black}{%
\fbox{\parbox{0.975\linewidth}{%
\textcolor{white!40!black}{\begin{tabular}{lr}%
  \begin{minipage}{0.6\textwidth}%
    {\small Copyright attribution to authors. \newline
    This work is a submission to SciPost Physics. \newline
    License information to appear upon publication. \newline
    Publication information to appear upon publication.}
  \end{minipage} & \begin{minipage}{0.4\textwidth}
    {\small Received Date \newline Accepted Date \newline Published Date}%
  \end{minipage}
\end{tabular}}
}}
}


\vspace{10pt}
\noindent\rule{\textwidth}{1pt}
\tableofcontents
\noindent\rule{\textwidth}{1pt}
\vspace{10pt}

\section{Introduction}
\label{sec:intro}

Higgs boson production in association with a bottom-quark pair ($pp\to\bbh$) is one of the 
processes contributing to inclusive Higgs production at the Large Hadron Collider (LHC). 
Although the inclusive $\bbh$ cross section is smaller by about two orders of magnitude~\cite{LHCHiggsCrossSectionWorkingGroup:2016ypw} compared to the main Higgs production channel which proceeds via gluon fusion,
the steadily increasing precision of LHC measurements keeps the $\bbh$ contribution phenomenologically relevant.
Unlike Higgs production in association with a top-quark pair ($t\bar{t}H$), which has an inclusive cross section of a similar order of magnitude to 
$\bbh$ production, the $\bbh$ final state has not yet been observed at the LHC. This is primarily due to the stringent 
$b$-jet identification requirements and the presence of overwhelming QCD backgrounds~\cite{Pagani:2020rsg}. As a result, 
$\bbh$ production is less sensitive for probing the bottom–Yukawa coupling compared to measurements based on the 
$H\to b\bar{b}$ decay channel. Nevertheless, a detailed study of $\bbh$ production remains important in the context of measuring the triple Higgs coupling~\cite{Manzoni:2023qaf}. 
In particular, $\bbh$ production constitutes one of the irreducible backgrounds to di-Higgs production when one of the Higgs bosons decays into a pair of bottom quarks. In addition, the 
$\bbh$ cross section can be enhanced in scenarios with more than one Higgs multiplet, 
such as in the Minimal Supersymmetric Standard Model (MSSM), due to modifications of the bottom–Yukawa coupling~\cite{Balazs:1998nt,Dawson:2005vi}. 

\begin{figure}[t!]
\centering
\begin{align}
	\sigma_{b;\lo}^{\mathrm{4FS}} & \sim \int d\mathrm{PS} \; \left|\;\includegraphics[width=1.9cm,trim={0 25mm 5mm 1mm}]{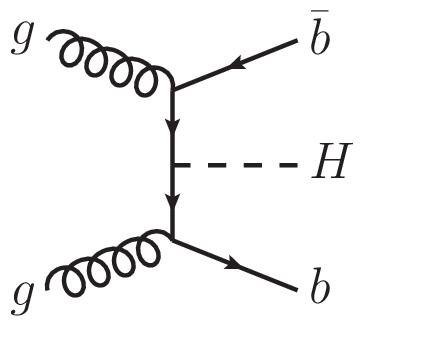} + \; \cdots \; \right|^2  
                                       \quad + \quad \int d\mathrm{PS} \; \left|\;\includegraphics[width=2.2cm,trim={0 20mm 1mm 0mm}]{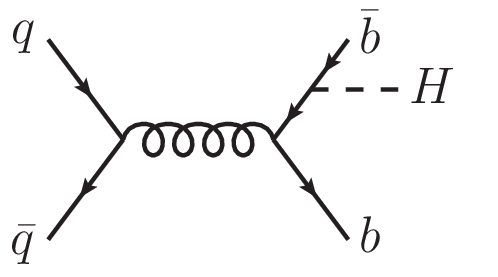} + \; \cdots \; \right|^2  \nonumber \\
	\sigma_{t;\lo}^{\mathrm{4FS}} & \sim \int d\mathrm{PS} \; \left|\;\includegraphics[width=2.4cm,trim={0 20mm 8mm 0mm}]{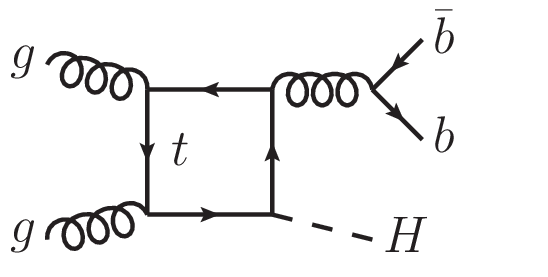} + \; \cdots \; \right|^2  
                                       \quad + \quad \int d\mathrm{PS} \; \left|\;\includegraphics[width=3cm,trim={0 20mm 1mm 0mm}]{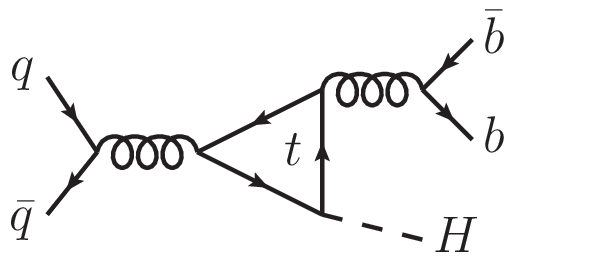} + \; \cdots \; \right|^2  \nonumber \\
	\sigma_{tb;\lo}^{\mathrm{4FS}} & \sim \int d\mathrm{PS} \;\; 2\mathrm{Re} \; 
	                                 \left(\;\includegraphics[width=1.9cm,trim={0 25mm 5mm 1mm}]{img/dsigma_b_lo_4FS_gg.eps} + \; \cdots \; \right)^*  
					 \; \left(\;\includegraphics[width=2.4cm,trim={0 20mm 8mm 0mm}]{img/dsigma_t_lo_4FS_gg.eps} + \; \cdots \; \right) \nonumber \\
					& \quad + \int d\mathrm{PS} \;\; 2\mathrm{Re} \;
	                                 \left(\;\includegraphics[width=2.2cm,trim={0 20mm 1mm 0mm}]{img/dsigma_b_lo_4FS_qq.eps} + \; \cdots \; \right)^*  
					 \; \left(\;\includegraphics[width=3cm,trim={0 20mm 1mm 0mm}]{img/dsigma_t_lo_4FS_qq.eps} + \; \cdots \; \right)  \nonumber
\end{align}
\caption{Representative Feynman diagrams illustrating the three $\bbh$ cross-section components defined in \cref{eq:SigmaBBHdecomposition} 
	at leading order in 4FS, grouped according to their dependence on the bottom- and top-Yukawa couplings.}
\label{fig:bbHxsec4FSfull}
\end{figure}

The inclusive cross section for $b\bar{b}H$ production receives contributions from terms proportional 
to the bottom- and top-quark Yukawa couplings, denoted by $y_b$ and $y_t$, respectively~\cite{Deutschmann:2018avk},
\begin{equation}
        \sigma_{\bbh} = y_b^2 \, \sigma_{b} + y_b y_t \,  \sigma_{tb}  +  y_t^2 \,  \sigma_{t} \, .
	\label{eq:SigmaBBHdecomposition}
\end{equation}
Here $\sigma_{b}$ denotes the contribution arising from the radiation of a Higgs boson off a bottom quark,
$\sigma_{t}$ corresponds to the contribution where the Higgs boson couples to a closed top-quark loop,
and $\sigma_{tb}$ represents the interference between these two Higgs emission mechanisms.~\footnote{The emission of a Higgs boson from 
a closed bottom-quark loop can also generate contributions proportional to $y_b$. These contributions are, however, subdominant. 
In the limit of a massless bottom quark, they vanish exactly.}
In \cref{fig:bbHxsec4FSfull} we show sample Feynman diagrams contributing to $\sigma_{b}$, $\sigma_{t}$ and $\sigma_{tb}$ at leading order in the four-flavour 
scheme (4FS), where the bottom quark is massive and excluded from the proton parton distribution functions (PDFs)\footnote{Similar decomposition in $y_b$ and $y_t$
can also be written for five-flavour scheme (5FS) calculation, 
where the bottom quark is massless and included in the proton PDFs. 
In the strict 5FS calculation, however, the $y_b y_t$ term vanishes.}. 
As can be seen at leading order, 
the $y_b$ contribution arises from tree-level amplitudes, whereas the $y_t$ contribution originates from one-loop amplitudes. Consequently, higher-order corrections to the $y_t$-type amplitudes are significantly more challenging to compute than those to the $y_b$-type ones.

\begin{figure}[t!]
\centering
\begin{align}
	\sigma_{t;\lo}^{\mathrm{4FS},\htl} & \sim \int d\mathrm{PS} \; \left|\;\includegraphics[width=2.4cm,trim={0 20mm 8mm 0mm}]{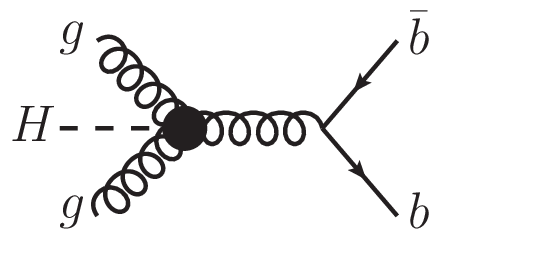} + \; \cdots \; \right|^2  
                                       \quad + \quad \int d\mathrm{PS} \; \left|\;\includegraphics[width=2.2cm,trim={0 24mm 1mm 0mm}]{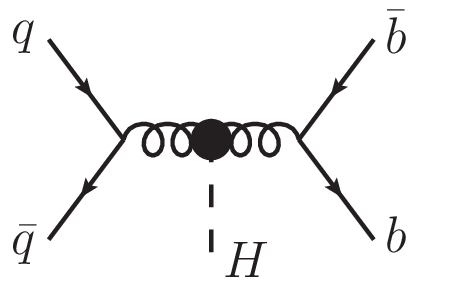} + \; \cdots \; \right|^2  \nonumber \\
	\sigma_{tb;\lo}^{\mathrm{4FS},\htl} & \sim \int d\mathrm{PS} \;\; 2\mathrm{Re} \; 
	                                 \left(\;\includegraphics[width=1.9cm,trim={0 25mm 5mm 1mm}]{img/dsigma_b_lo_4FS_gg.eps} + \; \cdots \; \right)^*  
					 \; \left(\;\includegraphics[width=2.4cm,trim={0 20mm 8mm 0mm}]{img/dsigma_t_lo_4FS_gg_HTL.eps} + \; \cdots \; \right) \nonumber \\
					& \quad + \int d\mathrm{PS} \;\; 2\mathrm{Re} \;
	                                 \left(\;\includegraphics[width=2.2cm,trim={0 20mm 1mm 0mm}]{img/dsigma_b_lo_4FS_qq.eps} + \; \cdots \; \right)^*  
					 \; \left(\;\includegraphics[width=2.2cm,trim={0 24mm 1mm 0mm}]{img/dsigma_t_lo_4FS_qq_HTL.eps} + \; \cdots \; \right)  \nonumber
\end{align}
\caption{Representative Feynman diagrams illustrating the $y_t$-dependent cross-section components defined in \cref{eq:SigmaBBHdecomposition} 
	at leading order in 4FS within the HTL framework.}
\label{fig:bbHxsec4FShtl}
\end{figure}

To facilitate the computation of the $y_t$ contribution, one can work in the heavy-top limit (HTL), where the top-quark mass is assumed to be much larger than the characteristic energy scale of the process. In this approximation, the top quark is integrated out and the Higgs–gluon interaction induced by the top-quark loop is described by a local effective operator~\cite{Wilczek:1977zn,Shifman:1978zn,Inami:1982xt}, which significantly simplifies the calculation of higher-order QCD corrections. The resulting effective Lagrangian takes the form
\begin{equation}
	\label{eq:HTLlagrangian}
	\mathcal{L} = -\frac{1}{4} \, C_1 \, H \, G_{\mu\nu}^{a}G^{a\mu\nu} \,,
\end{equation}
where $G_{\mu\nu}^a$ denotes the gluon field-strength tensor and
\begin{equation}
	C_1 = \frac{\as}{3\pi}\frac{y_t}{\sqrt{2} m_t} + \cO(\as^2) \,.
\end{equation}
In \cref{fig:bbHxsec4FShtl} we show sample Feynman diagrams entering the leading-order $\sigma_{t}$ and $\sigma_{tb}$ in the HTL.

Theoretical predictions for $\sigma_{b}$ have been computed up to NNLO QCD accuracy in 4FS~\cite{Dittmaier:2003ej,Dawson:2003kb,
Zhang:2017mdz,Biello:2024pgo}, 
while in 5FS computations up to N3LO QCD are available~\cite{Dicus:1998hs,Balazs:1998sb,Maltoni:2003pn,Harlander:2003ai,
Belyaev:2005bs,Harlander:2010cz,Ozeren:2010qp,Buhler:2012ytl,Harlander:2012pb,Harlander:2014hya,Ahmed:2014pka,Gehrmann:2014vha,AH:2019xds,
Duhr:2019kwi,Mondini:2021nck,Biello:2024vdh,Gavardi:2025zpf}.
Various approaches have been developed to reconcile the 5FS and 4FS $\sigma_b$ computations, with the aim of delivering reliable predictions throughout the full kinematic domain
~\cite{Harlander:2011aa,Bonvini:2015pxa,Forte:2015hba,Forte:2016sja,Duhr:2020kzd}.
The top-Yukawa induced cross sections, $\sigma_t$ and $\sigma_{tb}$, are known at NLO QCD in 4FS employing the Born-improved 
HTL framework~\cite{Deutschmann:2018avk,Manzoni:2023qaf}, while in 5FS they are estimated at LO using the $gg\to H$ inclusive
NNLO+PS generator~\cite{Hamilton:2012rf,Hamilton:2013fea,Hamilton:2015nsa}.
Among the three contributions, $\sigma_{tb}$ constitutes the smallest component, contributing at about -10\%~\cite{Dittmaier:2003ej,Deutschmann:2018avk}
while $\sigma_t$ is the dominant element with the cross section reaching twice as large as the $\sigma_b$~\cite{Deutschmann:2018avk,Manzoni:2023qaf,Biello:2025ksu}. 
The inclusion of NNLO QCD corrections in the state-of-the art
prediction of $\sigma_b$ was found to be crucial in bringing down the theoretical uncertainties due to scale variations to the level of 7-20\%, depending on the 
$b$-jet selections~\cite{Biello:2024pgo}. On the other hand, the most accurate computation of $\sigma_t$, which is at NLO QCD in HTL, 
still suffers from large theoretical uncertainties,
around 45\%~\cite{Deutschmann:2018avk,Manzoni:2023qaf,Biello:2025ksu}. 
It is therefore essential to improve the theoretical accuracy of $\sigma_t$ to reach NNLO QCD accuracy.

In this work we address the missing ingredient to obtain NNLO QCD prediction for $\sigma_t$ in the HTL effective theory framework in 4FS,
namely the two-loop amplitude. The two-loop scattering amplitude computation for $y_t$-induced $pp\to\bbh$ production with massive bottom quark is very challenging even using 
the most advanced multi-loop techniques. Following the approach taken in $\sigma_b$ NNLO QCD computation (matched to parton shower)~\cite{Biello:2024pgo},
the two-loop amplitude with massive bottom quark is constructed from the amplitude with massless bottom quark~\cite{Badger:2024mir} with the leading bottom-quark mass effects
included by exploiting universal factorisation properties of a massive amplitude~\cite{Mitov:2006xs}. Such an approach is often referred to as the \textit{massification} procedure.
While the universal, process independent factors encoding the leading mass effect are available, the process dependent two-loop massless amplitude for this process is not yet
available. The main result in the paper is the two-loop scattering amplitudes contributing to $y_t$-induced $pp\to\bbh$ production computed in HTL with 
the bottom quark is assumed to be massless.

The computation of two-loop five-particle scattering amplitude has received significant attention in the past few years, mainly due to the fact that it is
the main bottleneck in the computation of NNLO QCD corrections to a wide range of $2\to 3$ scattering processes analysed at the LHC.
Advancements in the Feynman integral computation via differential equations~\cite{Gehrmann:1999as,Henn:2013pwa}, 
particularly for the two-loop five point integrals~\cite{Gehrmann:2015bfy,Papadopoulos:2015jft,Gehrmann:2018yef,Abreu:2018rcw,Chicherin:2018mue,
Chicherin:2018old,Abreu:2018aqd,Abreu:2020jxa,Chicherin:2020oor,Canko:2020ylt,Chicherin:2021dyp,Abreu:2021smk,Kardos:2022tpo,Abreu:2023rco,Badger:2022hno,
Badger:2024fgb,FebresCordero:2023pww,Becchetti:2025oyb,Becchetti:2025qlu},
together with the adoption of finite-field techniques~\cite{vonManteuffel:2014ixa,Peraro:2016wsq,Klappert:2019emp,Peraro:2019svx,Smirnov:2019qkx,Klappert:2020aqs,Klappert:2020nbg}
as well as the development of highly optimised
integration-by-parts reduction strategies~\cite{Gluza:2010ws,Ita:2015tya,Larsen:2015ped,Wu:2023upw,Guan:2024byi}
have enabled the calculation of a large span of two-loop five-point amplitudes. Analytic results for fully massless two-loop five-point amplitudes are now
available in full colour~\cite{Badger:2019djh,Agarwal:2021vdh,Badger:2021imn,Abreu:2023bdp,
Badger:2023mgf,Agarwal:2023suw,DeLaurentis:2023nss,DeLaurentis:2023izi} while processes involving external and/or internal masses have been 
computed analytically in the leading colour approximation~\cite{Badger:2021nhg,Badger:2021ega,Abreu:2021asb,Badger:2022ncb,Badger:2024sqv,DeLaurentis:2025dxw,Badger:2025uym,
Badger:2025ljy},
except for the $y_b$-induced $pp\to\bbh$ amplitude~\cite{Badger:2024mir}.
Specifically for the two-loop five-point amplitude with one external mass, the class of amplitude that is the focus of this work, although the full set 
of Feynman integrals have been computed~\cite{Abreu:2023rco}, determining the rational coefficients appearing in the sub-leading colour amplitudes remains an extremely difficult task.
In this paper, we employ the leading colour approximation in order to keep the two-loop amplitude computation tractable.
The expected corrections to the matrix element from sub-leading colour terms are of $\mathcal{O}(N_c^{-2})$ in relative terms, i.e. about 10\% with respect to the leading colour matrix element. The actual impact on cross sections depends on the overall contribution arising from the two-loop matrix element that is unknown at this point. However, in Ref.~\cite{Czakon:2025wgs}, it has been demonstrated that this relatively simple estimate can be used to predict the impact of the missing terms on the level of differential cross sections for a range of $2 \to 3$ processes. Following a similar approach, the impact of the neglected terms can be traced as soon as NNLO QCD cross section results for $\sigma_t$ come available.

This paper is organised as follows. In \cref{sec:kinematicsetc}, we discuss the kinematical setup, partial amplitude decomposition and the construction of finite remainders.
In \cref{sec:framework}, our approach in deriving analytic form of the finite remainders using finite-field method from Feynman diagram input is explained.
In Section~\ref{sec:results}, we report benchmark numerical results at a physical phase-space point, outline the \textsc{C++} implementation, 
and investigate the numerical stability of the amplitude. Conclusions are given in Section~\ref{sec:conclusion}.

\section{Kinematics, colour decomposition and pole subtraction}
\label{sec:kinematicsetc}

We compute QCD corrections, up to two-loop level, to the following five-particle scattering processes contributing to top-Yukawa terms in the $\bbh$ production 
\begin{align}
	& 0 \to \bar{b}(p_1) + b(p_2) + g(p_3) + g(p_4) + H(p_5)\,, \\
	& 0 \to \bar{b}(p_1) + b(p_2) + \bar{q}(p_3) + q(p_4) + H(p_5)\,, \\
	& 0 \to \bar{b}(p_1) + b(p_2) + \bar{b}(p_3) + b(p_4) + H(p_5)\,.
	\label{eq:bbHprocesses}
\end{align}
We work in the heavy-top-quark limit (HTL) using effective interaction between Higgs boson and gluons as specified in \cref{eq:HTLlagrangian} 
and treat the bottom quark as massless. 
For completeness we also include the subprocesses contributing to bottom-induced $\bbh$ production, since the bottom quark is assumed to be massless.
All external momenta are taken to be outgoing and satisfy the on-shell conditions and momentum conservation 
\begin{align}
	& p_1^2 = p_2^2 = p_3^2 = p_4^2 = 0 \quad \mathrm{and} \quad  p_5^2=m_H^2 \,, \\
	& p_1 + p_2 + p_3 + p_4 + p_5 = 0 \,,
\end{align}
where $m_H$ denotes the Higgs boson mass. 
The scattering kinematics of $\bbh$ production is described by the following six scalar invariants
\begin{equation}
	\vec{s} = \lbrace s_{12}, s_{23}, s_{34}, s_{45}, s_{15}, m_H^2 \rbrace \,,
\end{equation}
with $s_{ij} = (p_i + p_j)^2$, together with the parity-odd invariant
\begin{equation}
	\tr_5 = 4i \varepsilon_{\mu\nu\rho\sigma} p_1^{\mu}p_2^{\nu}p_3^{\rho}p_4^{\sigma} \,,
\end{equation}
where $\varepsilon_{\mu\nu\rho\sigma}$ is the anti-symmetric Levi-Civita tensor.

\begin{figure}[t]
\centering
\includegraphics[width=14cm]{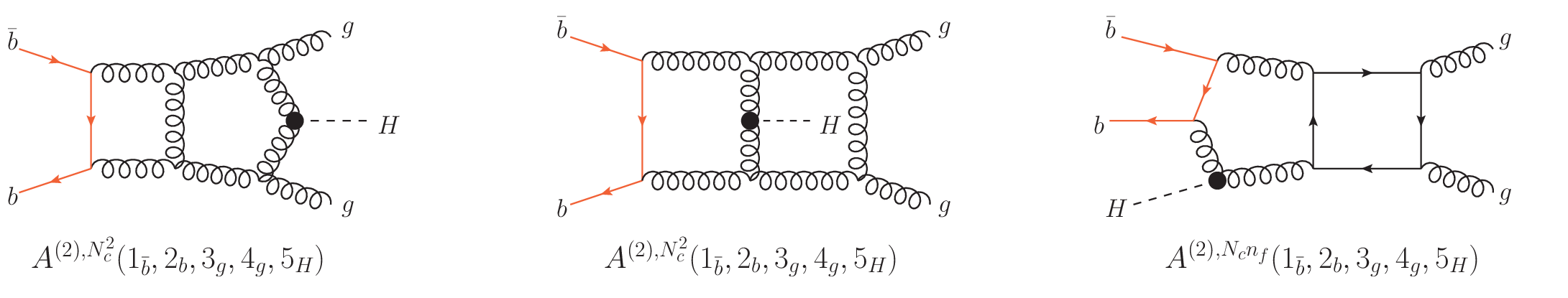}
\caption{Sample two-loop Feynman diagrams contributing to leading colour $\bbggH$ partial amplitudes.
	Solid lines denote massless quarks (bottom quarks in red), while dashed lines indicate the Higgs boson.}
\label{fig:ggdiagrams}
\end{figure}

\begin{figure}[t]
\centering
\includegraphics[width=14cm]{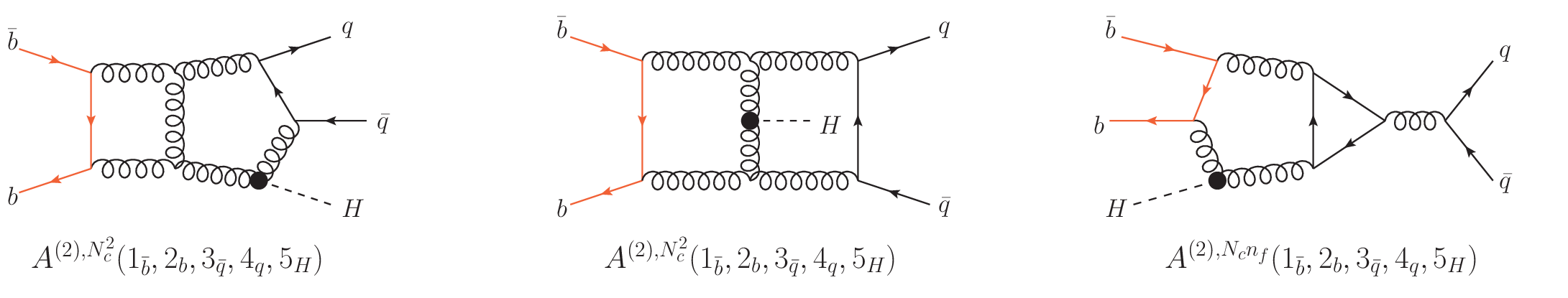}
\caption{Sample two-loop Feynman diagrams contributing to leading colour $\bbqqH$ partial amplitudes.
	Solid lines denote massless quarks (bottom quarks in red), while dashed lines indicate the Higgs boson.}
\label{fig:qqdiagrams}
\end{figure}

The $L$-loop scattering amplitudes $\cM^{(L)}$ are decomposed into partial amplitudes $\cA^{(L)}$ in the leading colour approximation according to
\begin{align}
	\cM^{(L)}_\lc\left(1_{\bar{b}},2_b,3_g,4_g,5_H\right) & = \bar{n}^L \bar{g}_s^2 \bar{C}_1 \; \bigg\lbrace 
	\left(t^{a_3}t^{a_4}\right)_{i_2}^{\;\bar{i}_1} \, \cA^{(L)}\left(1_{\bar{b}},2_b,3_g,4_g,5_H\right) + (3\leftrightarrow 4)
	\bigg\rbrace \,, 
        \label{eq:bbggHcolourdecomposition}	\\
	\cM^{(L)}_\lc\left(1_{\bar{b}},2_b,3_{\bar{q}},4_q,5_H\right) & = \bar{n}^L \bar{g}_s^2 \bar{C}_1 \;
	\delta_{i_4}^{\;\bar{i}_1}\delta_{i_2}^{\;\bar{i}_3} \; \cA^{(L)}\left(1_{\bar{b}},2_b,3_{\bar{q}},4_q,5_H\right) \,, 
        \label{eq:bbqqHcolourdecomposition}	\\
	\cM^{(L)}_\lc\left(1_{\bar{b}},2_b,3_{\bar{b}},4_b,5_H\right) & = \bar{n}^L \bar{g}_s^2 \bar{C}_1 \; \bigg\lbrace 
	  \delta_{i_4}^{\;\bar{i}_1}\delta_{i_2}^{\;\bar{i}_3} \; \cA^{(L)}\left(1_{\bar{b}},2_b,3_{\bar{q}},4_q,5_H\right) 
	  \label{eq:bbbbHcolourdecomposition} \\ 
	& \qquad\qquad\quad - \delta_{i_2}^{\;\bar{i}_1}\delta_{i_4}^{\;\bar{i}_3} \;  \cA^{(L)}\left(1_{\bar{b}},4_q,3_{\bar{q}},2_b,5_H\right) 
	\bigg\rbrace \,, \nonumber
\end{align}
where $t^a$'s are the $SU(N_c)$ fundamental generators satisfying $\tr(t^a t^b) = \delta^{ab}/2$, 
$\bar{g}_s$ is the bare strong coupling constant and $\bar C_1$ is the bare Higgs-gluon coupling.
The overall loop normalisation factor is
\begin{equation}
	\bar{n} =  (4\pi)^\eps e^{-\eps\gamma_E} \frac{\bar\as}{4\pi} \,.
\end{equation}
We see in \cref{eq:bbbbHcolourdecomposition} that the colour dressed $\bbbbH$ amplitude is made up completely by the $\bbqqH$ partial amplitude.
These partial amplitudes are further decomposed according to their dependence on the number of colour ($N_c$) and 
the number of closed massless quark loop ($n_f$). At leading colour the $(N_c,n_f)$ decomposition is given by
\begin{subequations}
\begin{align}
	\cA^{(1)} &= N_c \, A^{(1),N_c} + n_f \, A^{(1),n_f} \,,\\
	\cA^{(2)} &= N_c^2 \, A^{(2),N_c^2} + N_c n_f \, A^{(2),N_c n_f} + n_f^2 \, A^{(2),n_f^2}\,.
\end{align}
\label{eq:bareNcNfdecomposition}
\end{subequations}
We show in \cref{fig:ggdiagrams,fig:qqdiagrams} representative two-loop Feynman diagrams contributing to the leading colour $\bbggH$ and $\bbqqH$ 
partial amplitudes, respectively.


Ultra-violet (UV) singularities appearing in the one- and two-loop amplitudes are removed by adding the appropriate UV counterterms
due to the $\alpha_s$ and $C_1$~\cite{Harlander:2001is,Gehrmann:2011aa} renormalisations in the $\overline{\mathrm{MS}}$ scheme
\begin{align}
        \bar{\alpha}_s \mu_0^{2\eps} S_\eps & =  \alpha_s \mu_R^{2\eps} \, \bigg\lbrace 1 - \frac{\alpha_s}{4\pi} \frac{\beta_0}{\eps}
                                             + \left(\frac{\alpha_s}{4\pi}\right)^2 \left( \frac{\beta_0^2}{\eps^2} - \frac{\beta_1}{2\eps}\right) + \cO(\alpha_s^3) \bigg\rbrace \,, \\
	\bar{C}_1 & = C_1 \bigg[ 1 - \frac{\as}{4\pi} \frac{\beta_0}{\eps} 
	           + \bigg(\frac{\as}{4\pi}\bigg)^2 \bigg(\frac{\beta_0^2}{\eps^2} - \frac{\beta_1}{\eps}\bigg) \bigg] \,,
\end{align}
where $S_\eps = (4\pi)^\eps e^{-\eps \gamma_E}$ and $\mu_R$ is the renormalisation scale.
The renormalised partial amplitudes are then given by
\begin{subequations}
\begin{align}
	\cA^{(0)}_{\ren} & = \cA^{(0)} \,, \\
	\cA^{(1)}_{\ren} & = \cA^{(1)} - \frac{2\beta_0}{\eps}\cA^{(0)} \,, \\
	\cA^{(2)}_{\ren} & = \cA^{(2)} - \frac{3\beta_0}{\eps}\cA^{(1)} + \bigg( \frac{3\beta_0^2}{\eps^2} - \frac{3\beta_1}{2\eps} \bigg)\cA^{(0)} \,.
\end{align}
\label{eq:renormalisation}
\end{subequations}
The coefficients of $\beta$-function entering the UV renormalisation counterterms are
\begin{align}
\begin{aligned}
	\beta_0 &= \frac{11}{3}C_A - \frac{4}{3} T_F n_f \,, \\
	\beta_1 &= \frac{34}{3} C_A^2 - \frac{20}{3}C_A T_F n_f + 4 C_F n_f \,,
\end{aligned}
\end{align}
with 
\begin{equation}
	C_A = N_c, \qquad C_F = \frac{N_c}{2}, \qquad T_F = \frac{1}{2} \,.
\end{equation}
We note that $C_F$ is quoted in the leading colour limit.

The partial finite remainders up to two loops are obtained by subtracting the remaining infrared (IR) singularities from the renormalised 
partial amplitudes
\begin{subequations}
\begin{align}
	\cF^{(0)} & = \cA^{(0)}_{\ren} \,, \\
	\cF^{(1)} & = \cA^{(1)}_{\ren} - \zz^{(1)} \cA^{(0)}_{\ren} \,, \\
	\cF^{(2)} & = \cA^{(2)}_{\ren} - \zz^{(1)} \cA^{(1)}_{\ren} - \left[ \zz^{(2)} - (\zz^{(1)})^2 \right] \cA^{(0)}_{\ren} \,,
\end{align}
\label{eq:partialfinrem}
\end{subequations}
where $\zz^{(L)}$ is the $L$-loop universal IR poles~\cite{Catani:1998bh,Becher:2009cu,Gardi:2009qi,Becher:2009qa}. 
We adopt the IR poles subtraction construction of Refs.~\cite{Becher:2009cu,Becher:2009qa}, and the explicit expression for
$\zz^{(L)}$ up to two loops is given in ~\cref{app:IRcounterterms}.
Colour decomposition of the finite remainder for scattering processes contributing to $\bbh$ production follows that of the bare amplitude given in 
\cref{eq:bbggHcolourdecomposition,eq:bbqqHcolourdecomposition,eq:bbbbHcolourdecomposition}
\begin{align}
	\cR^{(L)}_\lc\left(1_{\bar{b}},2_b,3_g,4_g,5_H\right) & = n^L g_s^2 C_1 \; \bigg\lbrace 
	\left(t^{a_3}t^{a_4}\right)_{i_2}^{\;\bar{i}_1} \, \cF^{(L)}\left(1_{\bar{b}},2_b,3_g,4_g,5_H\right) + (3\leftrightarrow 4)
	\bigg\rbrace \,, \\
	\cR^{(L)}_\lc\left(1_{\bar{b}},2_b,3_{\bar{q}},4_q,5_H\right) & = n^L g_s^2 C_1 \;
	\delta_{i_4}^{\;\bar{i}_1}\delta_{i_2}^{\;\bar{i}_3} \; \cF^{(L)}\left(1_{\bar{b}},2_b,3_{\bar{q}},4_q,5_H\right) \,, \\
	\cR^{(L)}_\lc\left(1_{\bar{b}},2_b,3_{\bar{b}},4_b,5_H\right) & = n^L g_s^2 C_1 \; \bigg\lbrace 
	  \delta_{i_4}^{\;\bar{i}_1}\delta_{i_2}^{\;\bar{i}_3} \; \cF^{(L)}\left(1_{\bar{b}},2_b,3_{\bar{q}},4_q,5_H\right) \label{eq:partialdecomp} \\ 
	& \qquad\qquad\quad - \delta_{i_2}^{\;\bar{i}_1}\delta_{i_4}^{\;\bar{i}_3} \; \cF^{(L)}\left(1_{\bar{b}},4_q,3_{\bar{q}},2_b,5_H\right) 
	\bigg\rbrace \,, \nonumber
\end{align}
except for the $g_s$, $\alpha_s$ and $C_1$ couplings which are now renormalised quantities.
Furthermore, the finite remainders also admit the same $(N_c,n_f)$ decomposition in the leading colour approximation as for the bare amplitudes
\begin{subequations}
\begin{align}
	\cF^{(1)} &= N_c \, F^{(1),N_c} + n_f \, F^{(1),n_f} \,,\\
	\cF^{(2)} &= N_c^2 \, F^{(2),N_c^2} + N_c n_f \, F^{(2),N_c n_f} + n_f^2 \, F^{(2),n_f^2}\,.
\end{align}
\end{subequations}

We derive analytic expressions for the helicity-dependent leading colour partial finite remainders with the renormalisation scale $\mu_R$ set to 1.
The procedure to recover the $\mu_R$ dependence is discussed in ~\cref{app:mudep}. 
There are 8 (4) non-vanishing helicity configurations in the computation of $\bbggH$ ($\bbqqH$) scattering amplitudes.
We computed the following minimal set of helicity finite remainders
\begin{align}
	\bbggH: \quad & \cF^{(L)}\left(1^+_{\bar{b}},2^-_b,3^+_g,4^+_g,5_H\right)\,, \nonumber \\
	& \cF^{(L)}\left(1^+_{\bar{b}},2^-_b,3^+_g,4^-_g,5_H\right)\,, \label{eq:bbggHindephels}\\ 
	& \cF^{(L)}\left(1^-_{\bar{b}},2^+_b,3^+_g,4^-_g,5_H\right)\,,  \nonumber \\
	\bbqqH: \quad & \cF^{(L)}\left(1^+_{\bar{b}},2^-_b,3^+_{\bar{q}},4^-_q,5_H\right)\,,  \label{eq:bbqqHindephels} \\
	& \cF^{(L)}\left(1^+_{\bar{b}},2^-_b,3^-_{\bar{q}},4^+_q,5_H\right)\,, \nonumber
\end{align}
while the remaining helicity configurations are obtained by external momentum permutation and/or parity conjugation operation.

\section{Computational framework}
\label{sec:framework}

In this Section we describe the computational framework used to derive analytic expressions for the helicity-dependent finite remainders for the $\bbggH$
and $\bbqqH$ processes up to two loops. We generate the numerators of one- and two-loop leading-colour amplitudes from Feynman diagrams.
After the generation of relevant Feynman diagrams using \textsc{Qgraf}~\cite{Nogueira:1991ex}, we perform colour decomposition to extract independent set of bare partial amplitudes
which appear in ~\cref{eq:bbggHcolourdecomposition,eq:bbqqHcolourdecomposition,eq:bbbbHcolourdecomposition,eq:bareNcNfdecomposition}
and identify the loop integral topology for each diagram. We make use of the four-dimensional projection method~\cite{Peraro:2019cjj,Peraro:2020sfm} to cast the one- and two-loop partial amplitudes
as linear combinations of scalar Feynman integrals. The tensor decomposition for $\bbggH$ and $\bbqqH$ $y_t$-amplitudes considered in this paper is identical to that of 
$\bbggH$ and $\bbqqH$ $y_b$ contributions. We refer to Sections (3.1) and (3.2) of Ref.~\cite{Badger:2024mir} for a detailed discussion on the construction of $\bbggH$ and $\bbqqH$ helicity amplitudes 
using projectors in four dimensions. The use of projector method also allows us compute multiple helicity configurations simultaneously.
Algebraic manipulations done on the loop numerators are carried out using combinations of our in-house \textsc{Mathematica} and \textsc{Form}~\cite{Kuipers:2012rf,Ruijl:2017dtg,Davies:2026cci} scripts.

Scalar Feynman integrals appearing in the one- and two-loop partial amplitudes are expressed into a set of master integrals constructed in 
Refs.~\cite{Abreu:2020jxa,Canko:2020ylt,Abreu:2021smk,Kardos:2022tpo,Abreu:2023rco} by means
of integration-by-parts (IBP) reduction~\cite{Chetyrkin:1981qh,Tkachov:1981wb}. These master integrals are subsequently expressed in a basis of special functions 
called \textit{one-mass pentagon functions}~\cite{Gehrmann:2018yef,Chicherin:2020oor,Chicherin:2021dyp} followed by a 
Laurent expansion of the amplitude in the dimensional regulator parameter $\eps$ to the desired order ($\eps^2$ for one-loop amplitude and $\eps^0$ for the two-loop one).
At this stage the $L$-loop bare partial amplitude takes the form
\begin{align}
	A^{(L),h} = \sum_{i} \sum_{k=-2L}^{-2L+4}\, \eps^k \, c^{h}_{ik}(\lbrace p \rbrace) \, m_i(f) \;  + \; \cO(\eps^{-2L+5})\,, 
	\label{eq:bareampPfuncs}
\end{align}
where $L\in\lbrace 1,2\rbrace$, $\lbrace p \rbrace$ is a set of external momenta, $h$ the helicity configuration
and $m_i(f)$ is a monomial built out of one-mass pentagon functions and transcendental constants (e.g. $\pi^2$, $\zeta_3$).
Thanks to the deployment of pentagon function basis and Laurent expansion in $\eps$ we can explicitly subtract the UV and IR poles from the bare amplitude 
analytically to derive partial finite remainder according to \cref{eq:renormalisation,eq:partialfinrem}. After pole subtraction the finite remainder is expressed as
\begin{align}
	F^{(L),h} = \sum_{i} \sum_{k=0}^{-2L+4}\, \eps^k \, d^{h}_{ik}(\lbrace p \rbrace) \, m_i(f) \; + \; \cO(\eps^{-2L+5})\,,
\end{align}
for $L\in\lbrace 1,2\rbrace$.

Our analytic computation framework leverages the use of finite-field arithmetic to extract rational coefficients multiplying the pentagon function monomials
in order to sidestep large intermediate algebraic expressions typically encountered in the multi-scale loop scattering amplitude calculation.
To enable helicity amplitude computation with finite fields we use momentum twistor variables~\cite{Hodges:2009hk,Badger:2016uuq} 
to achieve rational parametrisation of external kinematics.
For the five-particle scattering with an external mass, the helicity amplitude can be expressed in terms of six momentum twistor variables~\cite{Badger:2021ega}
\begin{align}
	\vec{x} = \left\lbrace x_1, x_2, x_3, x_4, x_5, x_6 \right\rbrace\,.
\end{align}
We review the momentum twistor parametrisation for five-particle scattering with an external mass in \cref{app:momtwistor}.

The computation within the finite-field framework starts from unreduced bare partial amplitude representation and
all the algebraic manipulations that follow are replaced by finite-field numerical evaluation using \textsc{FiniteFlow}~\cite{Peraro:2016wsq,Peraro:2019svx} implementation. 
In particular, instead of deriving analytic IBP reduction table, optimized form of IBP 
relations~\cite{Gluza:2010ws,Schabinger:2011dz,Chen:2015lyz,Ita:2015tya,Larsen:2015ped,Zhang:2016kfo,Bohm:2017qme,Bosma:2018mtf,Boehm:2020zig}
that were generated using \textsc{NeatIBP} package~\cite{Wu:2023upw} are solved numerically
over finite fields and chained to the next step of the computation. We organize the IBP reduction by deriving optimized IBP relations only for one external momentum routing
of each integral families appearing in the IBP reduction. The solution for the full set of integral families (taking into account all the required external momentum permutations) 
are obtained numerically within the finite-field framework (see Appendix B of Ref.~\cite{Badger:2023xtl} for a more detailed discussion on IBP reduction approach that we adopt in this work). 

\renewcommand{\arraystretch}{1.9}
\begin{table}
    \centering
	\begin{tabular}{|cc|c|ccc|}
    \hline
		\multicolumn{3}{|c|}{}	&  $\cA^{(2),N_c^2}_{\bbggH}$ 	&  $\cA^{(2),N_c n_f}_{\bbggH}$  & $\cA^{(2),N_c^2}_{\bbqqH}$  \\
    \hline
		\multicolumn{3}{|c|}{\# of scalar integrals} & 139180 & 62337 & 54042  \\
    \hline
		PBmzz & \makecell{\includegraphics[width=2.2cm,trim={0 128mm 110mm 0mm},clip]{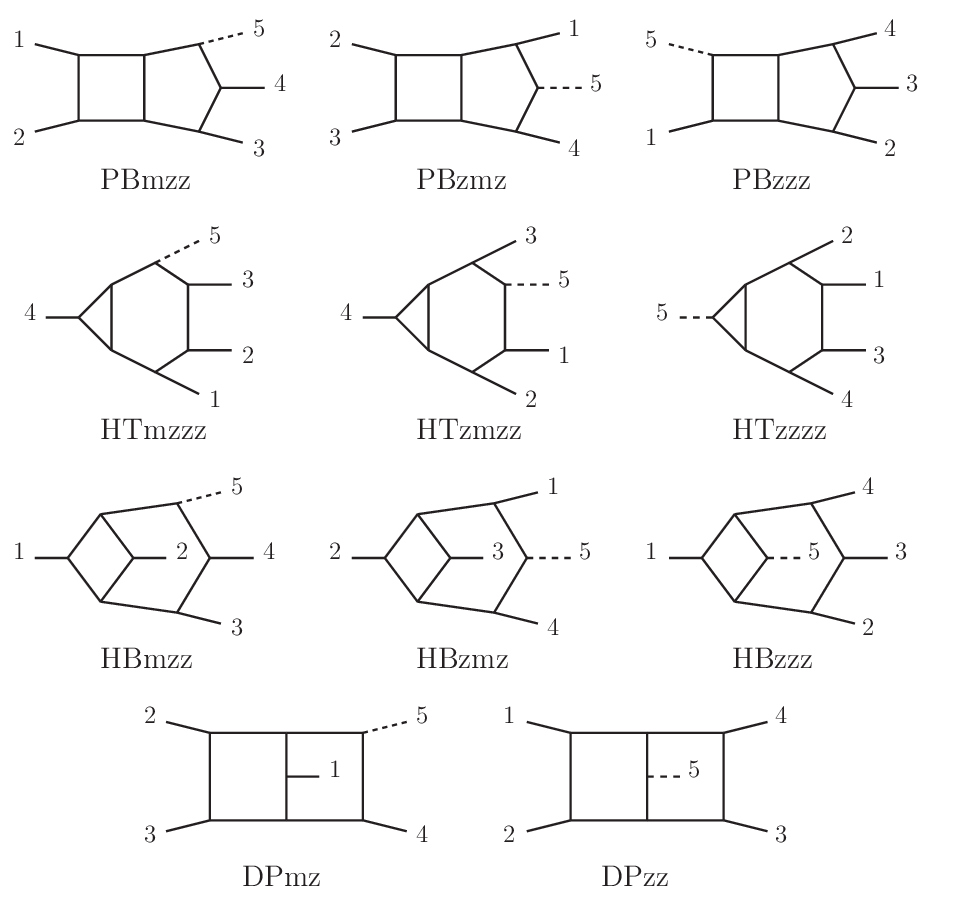}}      & $N_{\mathrm{perms}}$  & 6 & 4 & 4  \\
		PBzmz & \makecell{\includegraphics[width=2.2cm,trim={55mm 128mm 55.5mm 0mm},clip]{img/MainFam_Ordered.eps}}  & $N_{\mathrm{perms}}$  & 3 & 1 & 2  \\
		PBzzz & \makecell{\includegraphics[width=2.15cm,trim={109mm 128mm 3mm 0mm},clip]{img/MainFam_Ordered.eps}}   & $N_{\mathrm{perms}}$  & 6 & 6 & 4  \\
    \hline 
		HTmzz & \makecell{\includegraphics[width=2.15cm,trim={0 85mm 110mm 35mm},clip]{img/MainFam_Ordered.eps}}     & $N_{\mathrm{perms}}$  & 6 & 6 & 4  \\
		HTzmz & \makecell{\includegraphics[width=2.15cm,trim={55mm 85mm 55.5mm 35mm},clip]{img/MainFam_Ordered.eps}} & $N_{\mathrm{perms}}$  & 6 & 6 & 4  \\
		HTzzz & \makecell{\includegraphics[width=2.10cm,trim={109mm 85mm 4mm 35mm},clip]{img/MainFam_Ordered.eps}}   & $N_{\mathrm{perms}}$  & 4 & 3 & 4  \\
    \hline
		HBmzz & \makecell{\includegraphics[width=2.1cm,trim={0 47mm 110mm 75mm},clip]{img/MainFam_Ordered.eps}}      & $N_{\mathrm{perms}}$  & - & - & -  \\
		HBzmz & \makecell{\includegraphics[width=2.1cm,trim={55mm 47mm 55.5mm 75mm},clip]{img/MainFam_Ordered.eps}}  & $N_{\mathrm{perms}}$  & - & - & -  \\
		HBzzz & \makecell{\includegraphics[width=2.1cm,trim={109mm 47mm 4mm 75mm},clip]{img/MainFam_Ordered.eps}}    & $N_{\mathrm{perms}}$  & 4 & 4 & 4  \\
    \hline
		DPmz & \makecell{\includegraphics[width=2.25cm,trim={20mm 10mm 85mm 115mm},clip]{img/MainFam_Ordered.eps}}   & $N_{\mathrm{perms}}$  & - & - & -  \\
		DPzz & \makecell{\includegraphics[width=2.25cm,trim={80mm 10mm 25mm 115mm},clip]{img/MainFam_Ordered.eps}}   & $N_{\mathrm{perms}}$  & 2 & 1 & 2  \\
    \hline
    \end{tabular}
        \caption{ Integral families appearing in the two-loop five-point one-mass IBP reduction. The number of external momentum permutation for each family
	and the total number of scalar Feynman integrals prior to IBP reduction are presented 
	for the three most complicated partial amplitudes. As described in the text, 
	we take into account all the independent helicity configurations simultaneously. }
	\label{tab:integralfamilypermutation}
\end{table}

We show in \cref{tab:integralfamilypermutation} the number of external momentum permutation for each integral family appearing in the two-loop five-point computation with one-external mass,
for the three most complicated bare partial amplitudes: $A^{(2),N_c^2}$ and $A^{(2),N_c n_f}$ of $\bbggH$ and $A^{(2),N_c^2}$ of $\bbqqH$, with all independent helicity
configurations shown in \cref{eq:bbggHindephels,eq:bbqqHindephels} taken into account in a single computation.
To illustrate the complexity of the unreduced amplitudes, we also included in \cref{tab:integralfamilypermutation} the total number of scalar Feynman integrals to be
further processed by the IBP reduction. As we can see in \cref{tab:integralfamilypermutation}, although we are considering $\bbggH$
and $\bbqqH$ $y_t$-amplitudes in the leading colour limit, non-planar integral families (HBzzm and DPzz) are present in the computation, unlike the leading colour $\bbggH$
and $\bbqqH$ $y_b$-amplitudes. The presence of such non-planar families makes the present leading colour computation quite challenging, mainly due to the DPzz family
which IBP relation file size is more than an order of magnitude larger than the rest of the families~\cite{Badger:2024sqv,Badger:2024mir}.

\renewcommand{\arraystretch}{1.9}
\begin{table}[]
\centering
\begin{tabular}{|c|c|c|c|}
\hline
	& $\quad\cA^{(2),N_c^2}_{\bbggH}\quad$ & $\quad\cA^{(2),N_c n_f}_{\bbggH}\quad$ & $\quad\cA^{(2),N_c^2}_{\bbqqH}\quad$ \\
\hline
\hline
	$x_1 = 1$ & 256/251 & 166/161 & 145/140 \\
\hline
	\makecell{linear\\relations} & 187/183 & 157/153 & 118/114 \\
\hline
	\makecell{denominator\\matching} & 187/0 & 157/0 &  118/0 \\
\hline
	\makecell{univariate\\partial\\fraction in $x_5$}  & 53/52 & 50/49 & 41/38 \\
\hline
	\makecell{factor\\matching} & 53/0 & 50/0 & 41/0 \\
\hline
	\makecell{number of\\ sample points} & 157798 & 122037 & 60482\\
\hline
	\makecell{required \\ prime fields} & 3 & 2 & 3 \\
\hline
	\makecell{evaluation\\time per point}  & 2050s & 190s & 841s \\
\hline
\end{tabular}
	\caption{ Maximum polynomial degree in the numerator and denominator ($\mathrm{deg}_\mathrm{num}/\mathrm{deg}_\mathrm{den}$) 
	for the three most complicated bare partial amplitudes for different stages of rational coefficient optimisation. 
	The number of sample points (together with the required number of prime fields) 
	and evaluation time per sample point are also presented.
        As described in the text, we take into account all the independent helicity configurations simultaneously. 
	The evaluation time per point is measured on \textit{Intel(R) Xeon(R) Gold 6342 CPU @2.80GHz}.}
	\label{tab:reconstructiondata}
\end{table}

We extract analytic expressions of the rational coefficients of pentagon function monomials in the bare partial helicity amplitudes 
($c^h_{ik}$ in \cref{eq:bareampPfuncs}) from multiple numerical evaluations over finite fields. The complexity of such a reconstruction is
represented by the number of numerical sample points needed and evaluation time per sample point. In \cref{tab:reconstructiondata}
we present analytic reconstruction data for the three most complicated partial amplitudes considered in this work,
showing maximum polynomial degrees in the numerator and denominator, number of sample points required 
(including the number of prime fields needed to fully reconstruct the rational coefficients) and the evaluation time per sample point.
As is apparent in \cref{tab:reconstructiondata}, we have implemented a series of optimisation in order to reduce the complexity of rational coefficients
prior to performing analytic reconstruction.
These optimisations include: searching for linear relations among rational coefficients, 
recognising the denominator of rational coefficients using \textit{letters} appearing in the master integral computation~\cite{Abreu:2018zmy}
and performing univariate partial fraction decomposition before carrying out analytic reconstruction\footnote{We perform univariate 
partial fraction decomposition to the rational coefficients with respect to the $x_5$ variable.}.
We refer to Section 4 of Ref.~\cite{Badger:2021imn}, for the detailed description on the optimisations that we adopted in this work.
Once the analytic form of bare partial amplitudes are derived we perform UV and IR pole subtractions analytically according to \cref{eq:partialfinrem} to obtain the 
analytic representation of partial finite remainders. 
The final expressions for the finite remainders are further simplified using the \textsc{MultivariateApart} package~\cite{Heller:2021qkz}, 
resulting in a reduction of the output file size by a factor of approximately 1.2 to 4.

\section{Results}
\label{sec:results}

We have derived analytic results for the two-loop leading-colour amplitudes required in the computation of top-Yukawa contribution to $\bbh$ production at the LHC, 
within  the heavy-top quark mass effective theory framework. Explicit expressions for the finite remainders, together with the associated 
$\eps$-pole terms encoding both UV and IR singularities up to two loops, are provided in the ancillary material~\cite{zenodo} for all independent partial amplitudes and helicity configurations.
While the discussion in this paper is restricted to leading-colour amplitudes, the one-loop calculation has in fact been carried out retaining the full colour structure 
and the corresponding analytic expressions are also provided in full colour.

We have validated the results presented in this work by first comparing the full-colour one-loop amplitude 
against \textsc{OpenLoops}~\cite{Buccioni:2019sur} through $\mathcal{O}(\eps^0)$. The construction of the finite remainder also 
provides a strong check of the calculation, confirming the correctness of the singular structure of the 
amplitudes as predicted by universal UV and IR behaviour. 
The renormalisation-scale dependence of the finite remainder, as given in \cref{app:mudep}, is explicitly 
verified by performing numerical evaluations at a rescaled phase-space point. The resulting finite remainder, 
normalised to the mass dimension of the amplitude, agrees with the result obtained by including the $\mu_R$ dependence explicitly (see \cref{app:mudep}).
Finally, we perform a Ward identity check for the $\bbggH$ subprocess, demonstrating that the bare 
amplitude vanishes when one of the gluon polarisation vectors is replaced by its momentum.

To facilitate practical use, we supply \textsc{Mathematica} routines for the numerical evaluation of both the bare amplitudes and the corresponding hard functions. 
In addition, a dedicated \textsc{C++} library implementing the finite remainders and hard-function evaluations is included as part of the ancillary files~\cite{zenodo}. 
Detailed documentation of the \textsc{Mathematica} components is given in \cref{app:ancmath}, while instructions for compiling and using the \textsc{C++} library can be found in the \verb=anc_cpp/README= file.
In the following, we present representative benchmark results for the numerical evaluation of the hard functions across all partonic channels contributing to the 
$pp\to\bbh$ process, and we outline the structure and implementation of the C++ library.

\subsection*{Hard functions and benchmark evaluations}

The hard functions, built from the finite remainders, are defined up to two loops as
\begin{subequations}
\label{eq:hardfunctions}%
\begin{align}
\cH^{(0)} & = \colhelsum \big|\cR^{(0)}\big|^2\,, \\
\cH^{(1)} & = 2 \, \mathrm{Re} \, \colhelsum \cR^{(0)*} \cR^{(1)}\,, \\
\cH^{(2)} & = 2 \, \mathrm{Re} \,  \colhelsum \cR^{(0)*} \cR^{(2)}
             + \colhelsum \big|\cR^{(1)} \big|^2 \,,
\end{align}
\end{subequations}
where the overline corresponds to averaging over the initial-state colour and helicity configurations.
In evaluating the one- and two-loop hard functions, it suffices to retain the one-loop finite remainders up to $\cO(\eps^0)$.
We note that since the one-loop partial finite remainders $\cR^{(1)}$ are provided in full colour, in practice the two-loop hard function $\cH^{(2)}$
can be computed with $|\cR^{(1)}|^2$ either in the leading colour approximation or in full colour.

We list below the full set of partonic subprocesses contributing to $pp\to\bbh$ cross section computation 
\begin{subequations}
\label{eq:bbHchannels}
\begin{align}
        gg\to \bar{b}bH                   &:\qquad  g(-p_3) + g(-p_4) \to \bar{b}(p_1) + b(p_2) + H(p_5) \,, \\
        q\bar{q}\to \bar{b}bH             &:\qquad q(-p_3) + \bar{q}(-p_4) \to \bar{b}(p_1) + b(p_2) + H(p_5) \,, \\
        \bar{q}q\to \bar{b}bH             &:\qquad \bar{q}(-p_3) + q(-p_4) \to \bar{b}(p_1) + b(p_2) + H(p_5) \,, \\
        b\bar{b}\to \bar{b}bH             &:\qquad b(-p_3) + \bar{b}(-p_4) \to \bar{b}(p_1) + b(p_2) + H(p_5) \,, \\
        \bar{b}b\to \bar{b}bH             &:\qquad \bar{b}(-p_3) + b(-p_4)  \to \bar{b}(p_1) + b(p_2) + H(p_5) \,, \\
        bb\to bbH                         &:\qquad b(-p_3) + b(-p_4) \to b(p_1) + b(p_2) + H(p_5) \,, \\
        \bar{b}\bar{b}\to \bar{b}\bar{b}H &:\qquad \bar{b}(-p_3) + \bar{b}(-p_4) \to \bar{b}(p_1) + \bar{b}(p_2) + H(p_5) \,.
\end{align}
\end{subequations}
The evaluation of the hard functions for the channels listed above requires generating all partial finite remainders 
and helicity configurations from the analytically determined independent set through permutations of external momenta and/or parity transformations.
Performing such transformations on the analytic expressions written in terms of momentum twistor variables must be done for quantities free of spinor phase.
We choose the following spinor phases for $\bbggH$ and $\bbqqH$ processes, for the independent set of helicity configurations listed in \cref{eq:bbggHindephels,eq:bbqqHindephels}
\begin{align}
\label{eq:bbHphase}
\begin{alignedat}{5}
	\Phi_{\bbggh}^{+-++}    & = \frac{\spAB{2}{3}{1}}{\spA{3}{4}^2} \,, \qquad
	&&\Phi_{\bbggh}^{+-+-} && = \frac{\spAB{4}{1}{3}}{\spA{1}{3}\spB{2}{4}}  \,, \qquad 
	&&\Phi_{\bbggh}^{-++-} && = \frac{\spAB{4}{1}{3}}{\spA{2}{3}\spB{1}{4}}  \,, \\
	\Phi_{\bbqqh}^{+-+-}    & = \frac{\spA{2}{4}}{\spA{1}{3}} \,, \qquad
	&&\Phi_{\bbqqh}^{+--+} && = \frac{\spA{2}{3}}{\spA{1}{4}} \,.
	&& &&
\end{alignedat}
\end{align}
We refer to Section 4.1 of Ref.~\cite{Badger:2024mir} for the detailed discussion on the procedure to employ external momentum permutations and parity conjugation 
operation in the numerical evaluation of one-mass pentagon functions and the rational coefficients appearing in the finite remainder.

\renewcommand{\arraystretch}{1.5}
\begin{table}
    \centering
    \begin{tabular}{|c|ccc|}
    \hline
    $\bbggH$ & $\cH^{(0)}$ [GeV$^{-2}$] & $\cH^{(1)}/\cH^{(0)}$ & $\cH^{(2)}/\cH^{(0)}$  \\
    \hline
    $gg \to \bar{b}bH$ & $2.0771455 \cdot 10^{-9}$ & 0.23595251  & 0.132301423  \\
    \hline
    \hline
    $\bbqqH$ & $\cH^{(0)}$ [GeV$^{-2}$] & $\cH^{(1)}/\cH^{(0)}$ & $\cH^{(2)}/\cH^{(0)}$  \\
    \hline
    $q\bar{q} \to \bar{b}bH$ & $ 1.6517971 \cdot 10^{-9}$ &  -0.10509614 & 0.091278091 \\
    $\bar{q}q \to \bar{b}bH$ & $ 1.6517971 \cdot 10^{-9}$ &  -0.37281526 & 0.21389577 \\
    \hline
    \hline
    $\bbbbH$ & $\cH^{(0)}$ [GeV$^{-2}$] & $\cH^{(1)}/\cH^{(0)}$ & $\cH^{(2)}/\cH^{(0)}$  \\
    \hline
    $b\bar{b} \to \bar{b}bH$             & $ 4.0191158 \cdot 10^{-8}$ & 0.55822937 & 0.20414230 \\
    $\bar{b}b \to \bar{b}bH$             & $ 8.5181461 \cdot 10^{-9}$ & 0.33775653 & 0.17753505 \\
    $bb \to bbH$                         & $ 2.2702855 \cdot 10^{-8}$ & 0.54224045 & 0.11622574 \\
    $\bar{b}\bar{b} \to \bar{b}\bar{b}H$ & $ 2.2702855 \cdot 10^{-8}$ & 0.54224045 & 0.11622574 \\
    \hline
    \end{tabular}
	\caption{Numerical benchmarks for the hard functions of \cref{eq:hardfunctions}, 
	computed for the partonic processes listed in \cref{eq:bbHchannels} using the phase-space configuration and input parameters defined in 
	\cref{eq:PSpoint,eq:inputparameters}. We present the results of the hard functions in the strict leading colour limit where we keep only leading contributions in $N_c$ and $n_f$.}
    \label{tab:num-hardfunction}
\end{table}

In \cref{tab:num-hardfunction}, we present benchmark numerical results for the $y_t$ contribution to $pp\to\bbh$ scattering process up to two loops using the following
physical phase-space point
\begin{align}
\label{eq:PSpoint}
\begin{aligned}
p_1 & = ( 439.780219288, 162.506806585, 364.090479484, -185.570207307 ) \,, \\
p_2 & = ( 349.142399371, -17.5784710288, -333.450822239, 101.990007076 ) \,, \\
p_3 & = (-500,  0,  0,  -500)\,, \\
p_4 & = (-500,  0,  0,   500)\,, \\
p_5 & = ( 211.077381341, -144.928335556, -30.6396572450, 83.5802002309 )\,,
\end{aligned}
\end{align}
with the input parameters set to the values below:
\begin{equation}
        \label{eq:inputparameters}
	\mu_R = 173.2 \,, \qquad \alpha_s = 0.118 \,, \qquad \left(\frac{y_t}{\sqrt{2}m_t}\right)^{-1} = v = 246.220569073 \,,
\end{equation}
where $v$ is the Higgs vacuum expectation value. The momenta $p_i$, the renormalisation scale $\mu_R$ and $v$ are given in the units of GeV.
High-precision entries for the phase-space point in \cref{eq:PSpoint} are provided in the \textsc{Mathematica} ancillary files.
We present the results for the hard functions in the \textit{strict} leading colour limit where we only take into account
the leading terms in $N_c$ and $n_f$.

\subsection*{\textsc{C++} library: implementation and stability}

We implemented a \textsc{C++} library, which allows us to evaluate the partial amplitudes and assemble the hard functions in \cref{eq:hardfunctions} for arbitrary on-shell LHC kinematics across all relevant partonic channels. The software depends on the \textsc{PentagonFunctions++}~\cite{PentagonFunctions:cpp} implementation for evaluating pentagon functions and implements higher floating-point precision through the \textsc{QD} library~\cite{QDlib}, which provides \verb=dd_real= (32-digits) and \verb=qd_real= (64-digits) types. The software is provided as ancillary files with this article~\cite{zenodo} and has been tested to be consistent with the evaluation in \textsc{Mathematica}.

The software provides facilities to check numerical stability via a rescaling test. For a given phase space and parameter point $\{p_i,v,\mu_R\}$ this check re-evaluates the hard function for a scaled point $\{p_i',v',\mu_R'\} =  \{f \cdot p_i,f \cdot v,f \cdot \mu_R\}$ and compares the result with the expected scaling $\mathcal{H}^{(L)}(p_i',v',\mu_R') = f^{-2} \mathcal{H}^{(L)}(p_i,v,\mu_R)$. The number of agreeing digits provides an estimate for the actual numerical precision. A 'checked' evaluation routine is provided that aims to achieve a target precision by performing the rescaling test and, if necessary, increasing the floating-point precision. Heuristically, the main source of numerical instabilities lies in the rational functions, and therefore, floating-point precision is increased first. Only if their evaluation in \verb=qd_real= precision is not sufficient to reach the target precision, the precision of the pentagon functions is increased.

\begin{figure}
\centering
\includegraphics[width=0.9\textwidth]{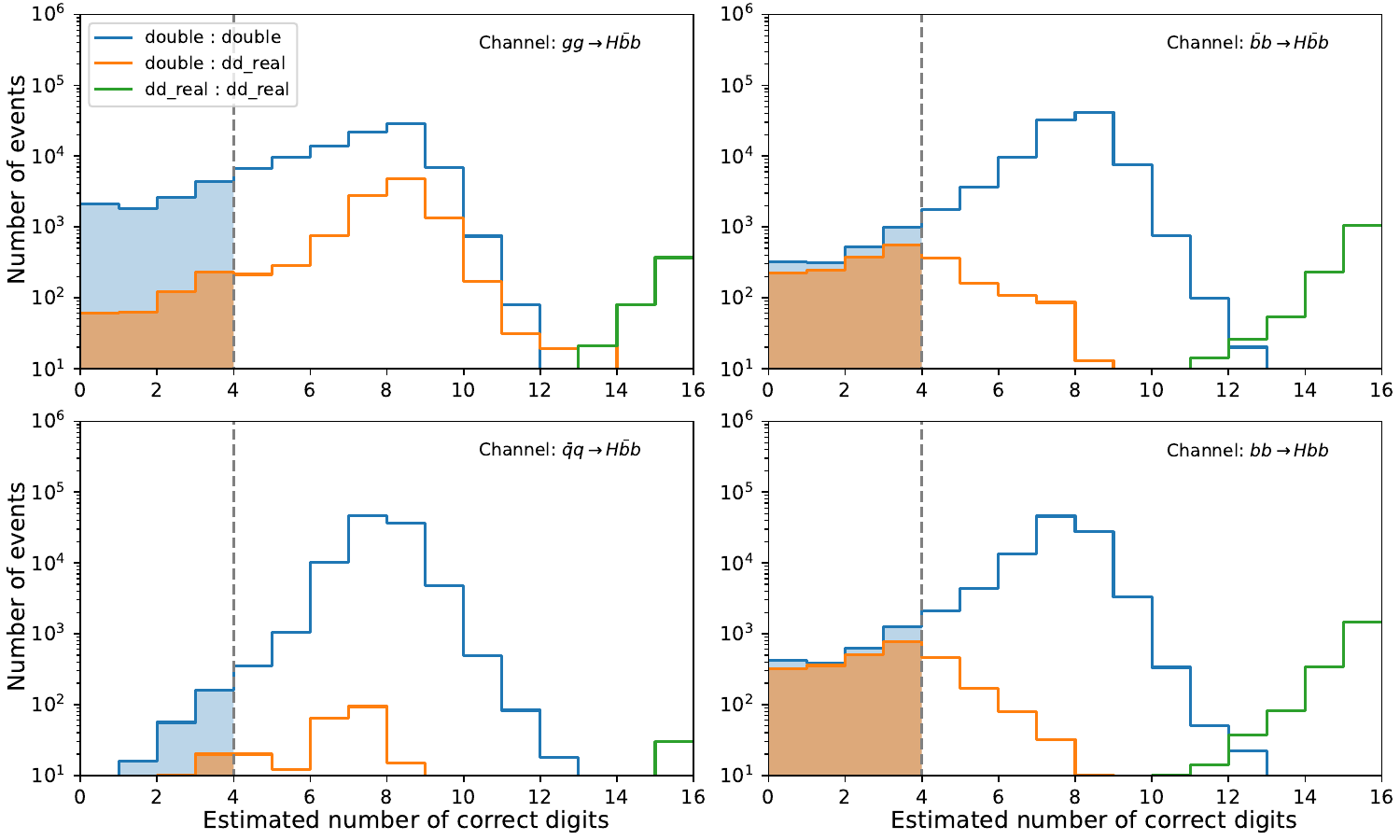}
	\cprotect\caption{The four panels show the distribution of the estimated numerical precision for four partonic channels ($gg\to\bar{b}bH$, $\bar{q}q\to\bar{b}bH$, 
	$\bar{b}b\to\bar{b}bH$ and $bb\to bbH$) based on 100k phase space points (the same for each channel). The blue line represents evaluations using double precision for the transcendental and rational parts. All points with fewer than 4 digits (indicated by the blue area) were re-evaluated with higher floating-point precision (\verb=dd_real=) for the rational part, shown as the orange line. Finally, all points that, even then, failed to achieve 4-digit precision (indicated by the orange area) were re-evaluated, also using \verb=dd_real= floats for the pentagon functions (the green line).}
\label{fig:stability}
\end{figure}

We tested this strategy and found it to be a good balance between speed and precision since the higher floating-point evaluation of the pentagon functions is costly. The results of these tests are summarized in \cref{fig:stability}. Overall, we find that about 10\% of the points in the $gg$ channel require a higher floating point treatment to achieve the target precision of 4-digits. A large fraction of these points can be rescued with the help of higher precision floats for the rational coefficients. Only 0.4\% of the points required higher-precision floats for the pentagon functions. On average, we found the evaluation time, including the test and potential rescue, to be about 8.5 seconds for the $gg$ on a simple \textit{Intel(R) Core(TM) i7-4790 CPU @3.60GHz}. For the $\bar{q}q$ channel we find 3 seconds evaluation time and a fraction of 0.2\% for higher rational floating point and 0.03\% for higher transcendental floating point evaluations.

The $gg$ channel is typically the most demanding, as it requires the most permutations of the partial amplitudes. A striking feature we noticed during the stability analysis, however, is that the $\bar{b}b$ and $bb$ channels behave quite differently from the $gg$ and $\bar{q}q$ channels in terms of stability. Here, the stability of the pentagon functions (and their permutations) was the limiting factor rather than the rational part. The main reason for this behaviour can be traced back to cancellations in the subtraction in \cref{eq:partialdecomp}. This leads (on average) to a higher evaluation time of about 13-15 seconds, since quite large fraction of events, 2\%, require evaluations with higher precision floats of the pentagon functions.

Additionally, we provide examples of how to use the library. First, we provide software, \textsc{eval} and \textsc{eval-checked}, that allows reproducing the numbers in \cref{tab:num-hardfunction} for the minimal set of partonic channels. The latter routine uses the aforementioned scaling check to ensure a target precision. Second, we provide two programs, \textsc{performance} and \textsc{stability}, to investigate the performance and numerical stability of the hard functions. These allow for reproducing the histograms and numbers above.

\section{Conclusion}
\label{sec:conclusion}

In this article we compute the two-loop five-point scattering amplitudes contributing to $pp\to\bbh$ production at the LHC
concentrating on the components induced by the top-quark Yukawa coupling. Analytic results for both $\bbggH$ and $\bbqqH$
are derived within the heavy-top-quark-limit (HTL) effective theory framework while treating the bottom quark as massless. To keep the 
computation tractable we have employed the leading colour approximation, where only leading terms in $N_c$ and $n_f$ are retained.
Our computation made use of a highly-refined framework that combines Feynman diagram approach, 
four-dimensional projection method, optimised IBP reduction strategy and analytic reconstruction techniques from finite-field evaluations. 
Analytic expressions for the finite remainders are written in terms of a one-mass pentagon-function monomial basis, 
together with the corresponding rational coefficients expressed in momentum twistor variables.
Our leading colour computation received contributions from non-planar Feynman integral families, which makes 
the determination of the rational coefficients significantly more challenging, compared to other typical leading-colour calculations.

Our work provides the missing ingredient for the computation of NNLO QCD $y_t$-induced $pp\to\bbh$ cross section within the HTL framework.
By supplementing our two-loop amplitudes, computed with massless bottom quarks, with leading bottom-quark mass effects via a massification procedure, 
4FS NNLO-accurate phenomenological studies can be performed. 
These can be carried out using techniques applicable to heavy-quark pair production in association with a colourless particle
either at fixed order (for example using sector-improved residue subtraction scheme (STRIPPER)~\cite{Czakon:2010td,Czakon:2014oma,Czakon:2019tmo} or
$q_T$-subtraction~\cite{Catani:2007vq,Bonciani:2015sha,Catani:2021cbl})
or within the \textsc{MiNNLOPS} framework~\cite{Biello:2024pgo} for matching to parton showers.
The \textsc{C++} numerical implementation provided in this work will immediately facilitates such phenomenological applications. 
Combined with the known $y_b$ contribution, our results will pave the way towards a full-fledged simulation of $pp\to\bbh$ production at NNLO QCD accuracy.

Although our two-loop amplitude computation is presented in the context of $b\bar{b}H$ production at the LHC, 
the scattering processes considered here also form essential building blocks for NNLO QCD predictions of 
$pp \to Hjj$ production within the HTL framework, in combination with the $H{+}4$ gluon channel.
The calculation of the two-loop $H{+}4$ gluon amplitudes is expected to be considerably more demanding, even at leading colour. 
This is due to the increased complexity of the unreduced amplitudes, the larger number of integral families and 
their associated momentum permutations entering the IBP reduction, as well as the likely appearance 
of rational functions of significantly higher degree than those encountered in the present work.
A significant increase in complexity is also anticipated when extending the current results to include subleading-colour contributions 
in the two-loop $y_t$-induced $\bbggH$ and $\bbqqH$ amplitudes.
Recent advances in handling rational coefficients within the finite-field framework, demonstrated in two-loop QCD computations for 
$pp \to jjj$~\cite{DeLaurentis:2023nss,DeLaurentis:2023izi} and $pp \to Vjj$~\cite{DeLaurentis:2025dxw}, 
may potentially offer advantages for the $H{+}4$-gluon channel 
as well as for full-colour $y_t$-induced $\bbggH$ and $\bbqqH$ calculations.

\section*{Acknowledgements}
We are grateful to Simon Badger, Colomba Brancaccio, and Simone Zoia for their collaboration 
during the initial stages of this project, and to Christian Biello for insightful discussions. 
We further thank Simon Badger and Colomba Brancaccio for their comments on the manuscript.

\paragraph{Funding information}
H.B.H. has been supported by an appointment to the JRG Program at the APCTP through the Science and Technology Promotion Fund and Lottery Fund 
of the Korean Government and by the Korean Local Governments – Gyeongsangbuk-do Province and Pohang City.
R.P. acknowledges that this research was funded in part by NCN 2024/55/D/ST2/00934.

\begin{appendix}
\numberwithin{equation}{section}

\section{Infrared counterterms}
\label{app:IRcounterterms}
This appendix provides explicit expressions for the IR poles relevant for the construction of partial finite remainder in \cref{eq:partialfinrem}.
Its perturbative expansion in the strong coupling constant is given by
\begin{equation}
	\zz = 1 + \frac{\as}{4\pi}\zz^{(1)} + \bigg(\frac{\as}{4\pi}\bigg)^2 \zz^{(2)} + \cO(\as^3) \,.
\end{equation}
The one-loop IR poles for $\bbggH$ process take the following form
\begin{align}
\begin{aligned}
	\zz_g^{(1),N_c} & = -\frac{3}{\eps^2} + \frac{1}{\eps} \bigg(-\frac{31}{6} + \ll_{14} + \ll_{23} + \ll_{34} - i\pi \bigg) \,, \\
	\zz_g^{(1),n_f} & = \frac{2}{3\eps} \,,
\end{aligned}
\end{align}
	where $\ell_{ij} = \log|s_{ij}|$. At two loops they read
\begin{align}
\begin{aligned}
	\zz_g^{(2),N_c^2} & = \frac{9}{2\eps^4} 
	                     + \frac{1}{\eps^3} \; \bigg(  3 i \pi - 3 \ll_{14} - 3 \ll_{23} - 3 \ll_{34} + \frac{95}{4}\bigg)  \\
			  & \quad +  \frac{1}{\eps^2} \; \bigg( 7 i \pi - 7 \ll_{14} - i \pi \ll_{14} - 7 \ll_{23} - i \pi \ll_{23} 
                                                     + \ll_{14} \ll_{23} - 7 \ll_{34} - i \pi \ll_{34} \\ 
				&	\qquad\qquad	     + \ll_{14} \ll_{34} + \ll_{23} \ll_{34} + \frac{1241}{72} - \frac{\pi^2}{4} 
				 		+ \frac{\ll_{14}^2}{2} + \frac{\ll_{23}^2}{2} + \frac{\ll_{34}^2}{2} \bigg) \\
			& \quad + \frac{1}{\eps} \; \bigg(
			-\frac{2513}{144} - \frac{67 i \pi}{18} + \frac{7 \pi^2}{72} + \frac{i \pi^3}{6} + \frac{9 \zeta_3}{2} + \frac{67 \ll_{14}}{18} \\
			& \qquad\qquad - \frac{\pi^2 \ll_{14}}{6} + \frac{67 \ll_{23}}{18} - \frac{\pi^2 \ll_{23}}{6} + \frac{67 \ll_{34}}{18} - \frac{\pi^2 \ll_{34}}{6}
			\bigg) \,, \\
	\zz_g^{(2),N_c n_f} & = -\frac{7}{2\eps^3} + \frac{1}{\eps^2}\;  \bigg( -i \pi + \ll_{14} + \ll_{23} + \ll_{34} - \frac{50}{9} \bigg) \\
	& \quad + \frac{1}{\eps} \; \bigg( \frac{125}{36} + \frac{5 i \pi}{9} + \frac{\pi^2}{36} - \frac{5 \ll_{14}}{9} - \frac{5 \ll_{23}}{9} - \frac{5 \ll_{34}}{9}\bigg) \,, \\
	\zz_g^{(2),n_f^2} & = \frac{4}{9\eps^2} \,.
\end{aligned}
\end{align}
For the $\bbqqH$ channel, the IR singularities at one-loop level are given by 
\begin{align}
\begin{aligned}
	\zz_q^{(1),N_c} & = \frac{2}{\eps^2} + \frac{1}{\eps}\;\bigg( -3 + \ll_{14} + \ll_{23}  \bigg) \,,\\
	\zz_q^{(1),n_f} & = 0 \,,
\end{aligned}
\end{align}
and at two loops they are
\begin{align}
\begin{aligned}
	\zz_q^{(2),N_c^2} & = \frac{2}{\eps^4} + \frac{1}{\eps^3}\;\bigg( -2 \ll_{14} - 2 \ll_{23} + \frac{23}{2} \bigg) \\
	                  & \quad + \frac{1}{\eps^2}\;\bigg( \ll_{14} \ll_{23} + \frac{113}{18} + \frac{\pi^2}{6} - \frac{29 \ll_{14}}{6} 
			    + \frac{\ll_{14}^2}{2} - \frac{29 \ll_{23}}{6} + \frac{\ll_{23}^2}{2} \bigg) \,, \\
			  & \quad + \frac{1}{\eps} \; \bigg( 7 \zeta_3 - \frac{2003}{216} - \frac{5 \pi^2}{12} + \frac{67 \ll_{14}}{18} 
			    - \frac{\pi^2 \ll_{14}}{6} + \frac{67 \ll_{23}}{18} - \frac{\pi^2 \ll_{23}}{6} \bigg) \,, \\
	\zz_q^{(2),N_c n_f} & = -\frac{1}{\eps^3} + \frac{1}{\eps^2} \; \bigg( -\frac{4}{9} + \frac{\ll_{14}}{3} + \frac{\ll_{23}}{3} \bigg) 
	                            + \frac{1}{\eps}\; \bigg( \frac{65}{54} + \frac{\pi^2}{6} - \frac{5 \ll_{14}}{9} - \frac{5 \ll_{23}}{9} \bigg) \,,\\
	\zz_q^{(2),n_f^2} & = 0 \,.
\end{aligned}
\end{align}

\section{Renormalisation scale dependence of the finite remainder}
\label{app:mudep}

The evaluation of partial finite remainder at an arbitrary value of $\mu_R$ from its expressions derived at $\mu_R=1$ requires
the additive $\mu_R$-restoring term $\delta F^{(L)}$
\begin{equation}
	F^{(L)}(\mu_R^2) = F^{(L)}(\mu_R^2=1) + \delta F^{(L)}(\mu_R^2) \,.
\end{equation}
For the $\bbggH$ partial finite remainders, the $\mu_R$-restoring terms are given by
\begin{align}
	\delta F_g^{(1),N_c}(\mu_R^2) & =  F_g^{(0)} \; \bigg\lbrace \left(\frac{13}{6} + g_1 \right) \, \log(\mu_R^2) 
	                                  + \frac{3}{2}\log^2(\mu_R^2) \bigg\rbrace\,, \\
	\delta F_g^{(1),n_f}(\mu_R^2) & = -\frac{2}{3}\log(\mu_R^2) \, F_g^{(0)}  \,, \\
	\delta F_g^{(2),N_c^2}(\mu_R^2) & =  F_g^{(1),N_c}(\mu_R^2=1) \; \bigg\lbrace \left(\frac{35}{6} + g_1 \right) \, \log(\mu_R^2) 
	                                  - \frac{3}{2}\log^2(\mu_R^2) \bigg\rbrace \\
					  & \quad + F_g^{(0)} \; \bigg\lbrace  \frac{9}{8}\log^4(\mu_R^2)
						+ \bigg( -\frac{61}{12} - \frac{3}{2}g_1 \bigg) \, \log^3(\mu_R^2) \nonumber \\
					  & \qquad\qquad\quad + \bigg( -\frac{349}{72} + 4 g_1 + \frac{g_1^2+\pi^2}{2}\bigg) \, \log^2(\mu_R^2) \nonumber \\
					  & \qquad\qquad\quad + \bigg( -\frac{65}{72} + 9\zeta_3 +\frac{67}{9}g_1 + \frac{7\pi^2}{36} - \frac{\pi^2}{3}g_1 \bigg) \, \log(\mu_R^2)
					  \bigg\rbrace \,, \nonumber \\
	\delta F_g^{(2),N_c n_f}(\mu_R^2) & = -\frac{4}{3} \log(\mu_R^2) \, F_g^{(1),N_c}(\mu_R^2=1) \\
				  & \quad  + F_g^{(1),n_f}(\mu_R^2=1) \; \bigg\lbrace \left(\frac{35}{6} + g_1 \right) \, \log(\mu_R^2) - \frac{3}{2}\log^2(\mu_R^2) \bigg\rbrace \nonumber \\
                                  & \quad + F_g^{(0)} \; \bigg\lbrace \frac{4}{3} \, \log^3(\mu_R^2) 
				                             + \bigg( -\frac{31}{8}-g_1 \bigg) \, \log^2(\mu_R^2) \nonumber \\
				& \qquad\qquad\quad + \bigg( -\frac{109}{18} -\frac{10}{9} g_1 + \frac{\pi^2}{18}\bigg) \, \log(\mu_R^2) \bigg\rbrace\,, \nonumber \\
	\delta F_g^{(2),n_f^2}(\mu_R^2) & = -\frac{4}{3} \log(\mu_R^2) \, F_g^{(1),n_f}(\mu_R^2=1) + \frac{4}{9}\log(\mu_R^2) \, F_g^{(0)} \,, 
\end{align}
while for the $\bbqqH$ process they take the following form
\begin{align}
        \delta F_q^{(1),N_c}(\mu_R^2) & =  F_q^{(0)} \; \bigg\lbrace \left(\frac{13}{3} + g_2 \right) \, \log(\mu_R^2)
                                          - \log^2(\mu_R^2) \bigg\rbrace\,, \\
	\delta F_q^{(1),n_f}(\mu_R^2) & = -\frac{4}{3}\log(\mu_R^2) \, F_q^{(0)}  \,, \\
	\delta F_q^{(2),N_c^2}(\mu_R^2) & =  F_q^{(1),N_c}(\mu_R^2=1) \; \bigg\lbrace \left( 8 + g_2 \right) \, \log(\mu_R^2)
                                          - \log^2(\mu_R^2) \bigg\rbrace \\
                                          & \quad + F_q^{(0)} \; \bigg\lbrace  \frac{1}{2}\log^4(\mu_R^2)
                                                + \bigg( -\frac{50}{9} - g_2 \bigg) \, \log^3(\mu_R^2) \nonumber \\
					  & \qquad\qquad\quad + \bigg( \frac{89}{9} + \frac{37}{6} g_2 + \frac{g_2^2}{2} + \frac{\pi^2}{3}\bigg) \, \log^2(\mu_R^2) \nonumber \\
					  & \qquad\qquad\quad + \bigg( \frac{1669}{108} +\frac{67}{9}g_2 - \bigg(\frac{5}{6} + \frac{g_2}{3}\bigg) \pi^2 
					                               + 14\zeta_3 \bigg) \, \log(\mu_R^2)
                                          \bigg\rbrace \,, \nonumber \\
        \delta F_q^{(2),N_c n_f}(\mu_R^2) & = -2 \log(\mu_R^2) \, F_q^{(1),N_c}(\mu_R^2=1) \\
                                  & \quad  + F_q^{(1),n_f}(\mu_R^2=1) \; \bigg\lbrace \left(8 + g_2 \right) \, \log(\mu_R^2) - \log^2(\mu_R^2) \bigg\rbrace \nonumber \\
                                  & \quad + F_q^{(0)} \; \bigg\lbrace \frac{14}{9} \, \log^3(\mu_R^2)
							     + \bigg( -\frac{77}{9}-\frac{5}{3}g_2 \bigg) \, \log^2(\mu_R^2) \nonumber \\
                                & \qquad\qquad\quad + \bigg( -\frac{286}{27} -\frac{10}{9} g_2 + \frac{\pi^2}{3}\bigg) \, \log(\mu_R^2) \bigg\rbrace\,, \nonumber \\
	\delta F_q^{(2),n_f^2}(\mu_R^2) & = -2 \log(\mu_R^2) \, F_q^{(1),n_f}(\mu_R^2=1) + \frac{4}{3}\log(\mu_R^2) \, F_q^{(0)} \,.
\end{align}
$g_1$ and $g_2$ are shorthands for the following linear combinations of one-mass pentagon functions
\begin{align}
	g_1 & = f_{1,2} + f_{1,5} + f_{1,7} -i\pi \,, \\
	g_2 & = f_{1,2} + f_{1,7}  \,. 
\end{align}
In terms of logarithms of Mandelstam invariants they read
\begin{align}
	g_1 & =  \ell_{14} + \ell_{23} +  \ell_{34}  -i\pi \,, \\
	g_2 & = \ell_{14} + \ell_{23}  \,. 
\end{align}

\section{Momentum twistor parametrisation}
\label{app:momtwistor}

In this Appendix we review the momentum twistor parametrisation for five-particle scattering in the presence of an external mass.
We start with momentum twistor parametrisation for massless six-particle scattering~\cite{Badger:2016uuq,Hartanto:2019uvl}
\begin{align}
	Z & = \begin{pmatrix}
	\lambda_1 & \lambda_2 & \lambda_3 & \lambda_4 & \lambda_5 & \lambda_6 \\
	\mu_1 & \mu_2 & \mu_3 & \mu_4 & \mu_5 & \mu_6 \\
        \end{pmatrix} \,, \\
	& = \begin{pmatrix}
	1 & 0 & \frac{1}{z_1} & \frac{1}{z_1} + \frac{1}{z_1 z_2} & \frac{1}{z_1} + \frac{1}{z_1 z_2} + \frac{1}{z_1 z_2 z_3} 
	& \frac{1}{z_1} + \frac{1}{z_1 z_2} + \frac{1}{z_1 z_2 z_3} + \frac{1}{z_1 z_2 z_3 z_4} \\
	0 & 1 & 1 & 1 & 1 & 1 \\
	0 & 0 & 0 & \frac{z_5}{z_2} & z_6 & 1 \\
	0 & 0 & 1 & 1 & z_7 & 1 - \frac{z_8}{z_5}
	\end{pmatrix} \,.
\end{align}
The holomorphic ($\lambda_i$) and anti-holomorphic $(\tilde\lambda_i)$ spinors for massless momenta $q_i$ in the 
massless six-particle scattering process can be explicitly written in terms of momentum twistor variable $z_i$ as follows
\begin{subequations}
\begin{alignat}{2}
	\lambda_{1\alpha} & = \begin{pmatrix} 
	1 \\ 0
	\end{pmatrix} \,,
	\qquad\qquad 
	&& \tilde\lambda_{1}^{\dot\alpha}  = \begin{pmatrix} 
		-1 +\frac{z_8}{z_5} \\ 1
	\end{pmatrix} \,, \\
	\lambda_{2\alpha} & = \begin{pmatrix} 
	0 \\ 1
	\end{pmatrix} \,,
	\qquad\qquad 
	&& \tilde\lambda_{2}^{\dot\alpha}  = \begin{pmatrix} 
		-z_1 \\ 0
	\end{pmatrix} \,, \\
	\lambda_{3\alpha} & = \begin{pmatrix} 
		\frac{1}{z_1} \\ 1
	\end{pmatrix} \,,
	\qquad\qquad 
	&& \tilde\lambda_{3}^{\dot\alpha} = \begin{pmatrix} 
		z_1 \\ z_1 z_5
	\end{pmatrix} \,, \\
	\lambda_{4\alpha} & = \begin{pmatrix} 
		\frac{1+z_2}{z_1 z_2} \\ 1
	\end{pmatrix} \,,
	\qquad\qquad 
	&& \tilde\lambda_{4}^{\dot\alpha} = \begin{pmatrix} 
		-z_1 z_2 z_3 (-1 + z_7) \\ -z_1 (z_5 + z_3 z_5 - z_2 z_3 z_6)
	\end{pmatrix} \,, \\
	\lambda_{5\alpha} & = \begin{pmatrix} 
		\frac{1+z_3+z_2 z_3}{z_1 z_2 z_3} \\ 1
	\end{pmatrix} \,,
	\qquad\qquad 
	&& \tilde\lambda_{5}^{\dot\alpha} = \begin{pmatrix} 
		\frac{z_1 z_2 z_3 \left( (1+z_4) z_5 (-1+z_7) + z_4 z_8 \right)}{z_5} \\ 
		-z_1 z_3 \left( -z_5 + z_2 ( z_4 ( -1 + z_6 ) + z_6 ) \right)
	\end{pmatrix} \,, \\
	\lambda_{6\alpha} & = \begin{pmatrix} 
		\frac{1 + z_4  + (1+z_2) z_3 z_4}{z_1 z_2 z_3 z_4} \\ 1
	\end{pmatrix} \,,
	\qquad\qquad 
	&& \tilde\lambda_{6}^{\dot\alpha} = \begin{pmatrix} 
		-\frac{ z_1 z_2 z_3 z_4 \left( z_5 (-1 + z_7) + z_8 \right)}{z_5} \\ z_1 z_2 z_3 z_4 (-1+z_6) 
	\end{pmatrix} \,.
\end{alignat}
\end{subequations}
We then assign the five-particle momenta $p_i$ from the six massless momenta $q_i$ as 
\begin{equation}
p_1 = q_1\,, \qquad
p_2 = q_2\,, \qquad
p_3 = q_3\,, \qquad
p_4 = q_4\,, \qquad
p_5 = q_5 + q_6 \,.
\end{equation}
To reduce the independent variables to six, we impose the following constraints
\begin{equation}
	\spA{q_2}{q_6} = 0 \qquad \mathrm{and} \qquad \spB{q_2}{q_6} = 0\,.
\end{equation}
These constraints in terms of six-particle momentum twistor variables read
\begin{align}
	-\frac{1+z_4 + z_3 z_4 + z_2 z_3 z_4 }{z_1 z_2 z_3 z_4} & = 0 \,, \\
	z_1^2 z_2  z_3 z_4 (-1 + z_6) & = 0 \,.
\end{align}
Solving for $z_4$ and $z_6$ and relabelling the momentum twistor variables as
\begin{equation}
z_1 \rightarrow x_1\,, \;
z_2 \rightarrow x_2\,, \;
z_3 \rightarrow x_3\,, \;
z_5 \rightarrow x_4\,, \;
z_7 \rightarrow x_5\,, \;
z_8 \rightarrow x_6\,, \;
\end{equation}
the holomorphic and anti-holomorphic spinors for the massless momenta $p_i$ in the five-particle scattering with an external mass take the form
\begin{subequations}
\begin{alignat}{2}
	\lambda_{1\alpha} & = \begin{pmatrix} 
	1 \\ 0
	\end{pmatrix} \,,
	\qquad\qquad 
	&& \tilde\lambda_{1}^{\dot\alpha}  = \begin{pmatrix} 
		-1 +\frac{x_6}{x_4} \\ 1
	\end{pmatrix} \,, \\
	\lambda_{2\alpha} & = \begin{pmatrix} 
	0 \\ 1
	\end{pmatrix} \,,
	\qquad\qquad 
	&& \tilde\lambda_{2}^{\dot\alpha}  = \begin{pmatrix} 
		-x_1 \\ 0
	\end{pmatrix} \,, \\
	\lambda_{3\alpha} & = \begin{pmatrix} 
		\frac{1}{x_1} \\ 1
	\end{pmatrix} \,,
	\qquad\qquad 
	&& \tilde\lambda_{3}^{\dot\alpha} = \begin{pmatrix} 
		x_1 \\ x_1 x_4
	\end{pmatrix} \,, \\
	\lambda_{4\alpha} & = \begin{pmatrix} 
		\frac{1+x_2}{x_1 x_2} \\ 1
	\end{pmatrix} \,,
	\qquad\qquad 
	&& \tilde\lambda_{4}^{\dot\alpha} = \begin{pmatrix} 
		-x_1 x_2 x_3 (-1 + x_5) \\ x_1 ( x_2 x_3 - (1+x_3) x_4 )
	\end{pmatrix} \,, 
\end{alignat}
\label{eq:spinors5p1m}
\end{subequations}
and for the Higgs-boson momentum we have 
\begin{equation}
	p_{5\alpha \dot\alpha} = \begin{pmatrix}
		-\frac{ (1+(1+x_2) x_3) (x_2-x_4) }{x_2} & -\left( (1+x_2) x_3 (-1+x_5) \right) + \frac{x_6}{x_4} \\
		x_1 x_3 (-x_2+x_4) & -x_1 x_2 x_3 (-1+x_5)
	\end{pmatrix} \,.
	\label{eq:HiggsMomentum}
\end{equation}
Using \cref{eq:spinors5p1m,eq:HiggsMomentum} we can write any scalar products and spinor strings built out of external momenta in terms 
of five-point momentum twistor variable $x_i$.
Inversely, the five-point momentum twistor variables $x_i$ can be written in terms of scalar invariants and spinor products of external momenta
\begin{equation}
\begin{alignedat}{2}
  & x_1 = s_{12} \,, \qquad && x_2 = -\frac{\trp(1234)}{s_{12}s_{34}} \,, \\
	& x_3 = \frac{\trp(134152)}{s_{13}\trp{(1452)}} \,, \qquad && x_4 = \frac{s_{23}}{s_{12}} \,, \\
  & x_5 = -\frac{\trm(1(2+3)(1+5)523)}{s_{23}\trm(1523)} \,, \qquad && x_6 = \frac{s_{123}}{s_{12}} \,,
\end{alignedat}
\label{eq:momtwistor5pt}
\end{equation}
where
\begin{align}
& \tr_{\pm}(ij \cdots kl) = \frac{1}{2} \tr\left[(1\pm\gamma_5)\slashed{p}_i\slashed{p}_j \cdots\slashed{p}_k\slashed{p}_l\right] \,, \\
	& \tr_{\pm}(\cdots (i+j) \cdots) = \tr_{\pm}(\cdots i \cdots) + \tr_{\pm}(\cdots j \cdots) \,.
\end{align}
In practice, in the numerical evaluation, the momentum twistor variables are evaluated from the four momenta using \cref{eq:momtwistor5pt}.

\section{Overview of the \textsc{Mathematica} ancillary files}
\label{app:ancmath}

This appendix provides an overview of the \textsc{Mathematica} ancillary material accompanying this work, available from~\cite{zenodo}. 
Documentation for the associated \textsc{C++} library is instead included in the file \verb=anc_cpp/README= within the same repository~\cite{zenodo}.
The analytic results are organised by separating the finite remainders from the $\eps$-pole contributions, the latter encoding both ultraviolet and infrared singularities. 
The $L$-loop finite remainders, denoted by $F^{(L)}$, are expressed in the form
\begin{equation}
\label{eq:finremsparse}
        F^{(L)} = \sum_{p=0}^{-2L+4} \eps^p \, \sum_{ij} r_{i}(\vec{y}) \, S_{pij} \, m_j(f) \,,
        \qquad L \in \{1,2\} \,.
\end{equation}
Here, $r_i$ denotes independent rational coefficients and $S$ is a sparse rational matrix. 
The monomials $m_j(f)$ are constructed from pentagon functions, transcendental constants, square roots, and powers of $x_1$, included to ensure that $m_j(f)$ is dimensionless. 
The coefficients $r_i(\vec{y})$ in \cref{eq:finremsparse} are expressed in terms of polynomial factors $y_i(\vec{x})$, 
collected across finite remainders with the same loop order and helicity configurations.
The analytic results are derived at $x_1 = 1$; its dependence is explicit in $m_j(f)$ and restored for the coefficients through the numerical evaluation routines.

All analytic results presented in this work are made available in a \textsc{Mathematica} format. 
In what follows, we detail the correspondence between the notation 
used in the ancillary material and that adopted in the main text.
\begin{itemize}
\item Partonic subprocesses contributing to $y_t$-terms in $pp\to b\bar{b}H$ scattering:
\begin{align}
\label{eq:translateproc}
\begin{aligned}
        \texttt{bbggH\_yt} &= \bbggH \,, \\
        \texttt{bbqqH\_yt} &= \bbqqH \,.
\end{aligned}
\end{align}
\item Colour factors in $\bbggH$:
\begin{align}
\label{eq:translatecolgg}
\begin{aligned}
        \texttt{T2341}  &= (t^{a_3}t^{a_4})_{i_2}^{\;\;\bar{i}_1} \,, \\
        \texttt{d12d34} &= \delta_{i_2}^{\;\;\bar{i}_1} \delta^{a_3 a_4} \,.
\end{aligned}
\end{align}
The colour factor $\delta_{i_2}^{\;\;\bar{i}_1} \delta^{a_3 a_4}$ is a subleading contribution in the $\bbggH$ colour decomposition and 
included only in the one-loop amplitudes.
\item Colour factors in $\bbqqH$:
\begin{align}
\label{eq:translatecolqq}
\begin{aligned}
        \texttt{d14d23} &= \delta_{i_4}^{\;\;\bar{i}_1}  \delta_{i_2}^{\;\;\bar{i}_3}  \,,  \\
        \texttt{d12d34} &= \delta_{i_2}^{\;\;\bar{i}_1}  \delta_{i_4}^{\;\;\bar{i}_3}  \,.
\end{aligned}
\end{align}
The colour factor $\delta_{i_2}^{\;\;\bar{i}_1} \delta_{i_4}^{\;\;\bar{i}_3}$ is a subleading contribution in the $\bbqqH$ colour decomposition and 
included only in the one-loop amplitudes.
\item Helicity configurations:
\begin{align}
\label{eq:translatehel}
\begin{aligned}
        \texttt{pmpp} &= +-++ \,, \\
        \texttt{pmpm} &= +-+- \,, \\
        \texttt{pmmp} &= +--+ \,, \\
        \texttt{mppm} &= -++- \,. \\
\end{aligned}
\end{align}
\item Dimensional regulator, momentum twistor variables, Mandelstam invariants and other kinematical quantities needed to evaluate the amplitude:
\begin{align}
\label{eq:translatekinematics}
\begin{aligned}
    \texttt{eps} & = \eps \,, \\
        \texttt{ex[i]} &= x_i \quad \forall i = 1,\dots,6 \,, \\
        \texttt{s[i,j]} \; \mathrm{or} \; \texttt{sij} &= s_{ij} \,, \\
        \texttt{s[i,j,k]}  \; \mathrm{or} \; \texttt{sijk} &= s_{ijk} \,, \\
        \texttt{tr5} & = \tr_5 \,, \\
        \texttt{trp[i,j,k,l]} & = \tr_+(ijkl) \,, \\
        \texttt{trm[i,j,k,l]} & = \tr_-(ijkl) \,. \\
\end{aligned}
\end{align}
\item Pentagon functions, transcendental constants and square roots (see Refs.~\cite{Abreu:2023rco,Chicherin:2021dyp} for a more detailed description of these objects): 
\begin{align}
\label{eq:translatepfuncs}
\begin{aligned}
	\texttt{F[w,i]} &= f_{w,i} \,, \\
        \texttt{re[3,1]} &= \zeta_3 \,, \\
        \texttt{im[1,1]} & = \i \pi \,, \\
        \texttt{sqrtDelta5} & = \sqrt{\Delta_5} \,, \\
        \texttt{sqrtG3[i]} & = \sqrt{\Delta_3^{(i)}} \,, \\
        \texttt{sqrtSigma5[i]} & = \sqrt{\Sigma_5^{(i)}} \,.
\end{aligned}
\end{align}
\item Renormalisation scale:
\begin{align}
\label{eq:translatemu}
\begin{aligned}
        \texttt{mu} &= \mu_R \,. \\
\end{aligned}
\end{align}
\end{itemize}

The ancillary files are organised as follows:
\begin{itemize}
\item scattering channel \verb=<proc>=  (\verb=bbggH_yt=, \verb=bbqqH_yt=),
\item loop order \verb=<loop>= (\verb=1=, \verb=2=),
\item colour factor \verb=<col>=: (\verb=T2341= and  \verb=d12d34= for $\bbggH$, \verb=d14d23= and  \verb=d12d34= for $\bbqqH$),
\item powers of $n_f$ (\verb=<a>=) and $N_c$ (\verb=<b>=),
\item helicity configuration \verb=<hel>=: (\verb=pmpp=, \verb=pmpm=, \verb=pmmp=, \verb=mppm=).
\end{itemize}
In the following, we list and describe all files included.
\begin{itemize}

        \item \verb=amplitudes_<proc>/tree_<proc>_<col>_Nfp<a>_Ncp<b>_<hel>.m=:
                tree-level helicity amplitude.
        \item \verb=amplitudes_<proc>/finrem_coeffs_y_<loop>L_<proc>_<col>_Nfp<a>_Ncp<b>_<hel>.m=:
        independent rational coefficients $r_i$ in \cref{eq:finremsparse} as functions of common factors \verb=y[i]= ($y_i$).

        \item \verb=amplitudes_<proc>/finrem_sm_<loop>L_<proc>_<col>_Nfp<a>_Ncp<b>_<hel>.m=:
                rational matrix $S$ in \cref{eq:finremsparse}, written in \textsc{Mathematica}'s \verb=SparseArray= format.

        \item \verb=amplitudes_<proc>/ys_<loop>L_<proc>_<hel>.m=: replacement rules for the common polynomial factors \verb=y[i]=  ($y_i$) in
                terms of momentum twistor variables \verb=ex[i]=  ($x_i$).

        \item \verb=amplitudes_<proc>/FunctionBasis_<proc>_<loop>L.m=: monomial basis of pentagon functions, transcendental constants and square roots
                ($m_j(f)$ in \cref{eq:finremsparse}).

        \item \verb=amplitudes_<proc>/mudep_<loop>L_<proc>_<col>_Nfp<a>_Ncp<b>.m=: $\mu_R$-restoring term as explicitly given in \cref{app:mudep}.

	\item \verb=amplitudes_<proc>/poles_<loop>L_<proc>_<col>_Nfp<a>_Ncp<b>_<hel>.m=: pole contributions containing UV and IR counterterms, 
		taking the same form as the bare helicity amplitudes in \cref{eq:bareampPfuncs}.
                The pole terms are given in the following format,
        \begin{center}
        \verb={coefficientrules,{coefficients,monomials}}= \,,
        \end{center}
        where \verb=coefficientrules= is a list of replacement rules defining the independent rational coefficients \verb=f[i]=
        in terms of the momentum twistor variables \verb=ex[i]=,
        \verb=monomials= and \verb=coefficients= are the lists of one-mass pentagon-function monomials and
                 of their coefficients, which depend on both \verb=eps= ($\eps$) and the linearly independent rational coefficients \verb=f[i]=.

        \item \verb=F_permutations_sm/=: a folder containing the pentagon functions' permutation rules. The permutation to the ordering $\texttt{<perm>}$ of the external momenta is given by the matrix stored in \verb=sm_F_perm_2l5p1m_<perm>.m= (in \textsc{Mathematica}'s \verb=SparseArray=
                format). Multiplying the latter by the array of pentagon-function monomials in
                \verb=F_monomials.m= gives the permutation of the pentagon functions listed in the variable \verb=listOfFunctions= in \verb=utilities.m=.

        \item \verb=Evaluate_BareAmplitudes_<proc>.wl=: a script to numerically evaluate the bare helicity partial 
		amplitudes for the independent helicity configurations in the $gg\to\bar{b}bH$ and $\bar{q}q \to\bar{b}bH$ scattering channels
                as defined in \cref{eq:bbHchannels}.

        \item \verb=Evaluate_HardFunctions_<proc>.wl=: script to evaluate numerically the hard functions for all partonic channels
		contributing to $pp\to b\bar{b}H$ as listed in \cref{eq:bbHchannels}. The evaluation of 
		two-loop hard function $\cH^{(2)}$ in \cref{eq:hardfunctions} is provided with $|\cR^{(1)}|$ computed both in leading and full colour.

        \item \verb=utilities.m=: a collection of auxiliary functions needed for the numerical evaluation scripts.

\end{itemize}

\end{appendix}


\bibliography{bbh_yt_lc.bib}

\end{document}